\documentclass[11pt,a4paper]{article}
\pdfoutput=1 
\usepackage[dvipdfmx]{graphicx}
\usepackage[dvipdfmx]{color} 

\usepackage{amsmath,amssymb,amsfonts,color,scalefnt}
\usepackage{epsfig,amsthm}
\usepackage{soul}
\usepackage[makeroom]{cancel}
\usepackage{mathrsfs}
\usepackage{epstopdf}
\usepackage[utf8]{inputenc}
\usepackage{hyperref}
\usepackage{slashed}
\usepackage{pict2e}
\usepackage{tikz}
\usepackage{cite}
\usepackage{float}
\usepackage[percent]{overpic}
\usepackage{multirow}
\usepackage{bbold}
\usepackage{physics}
\usepackage{booktabs}
\usepackage{mwe}
\usepackage [autostyle, english = american]{csquotes}
\usepackage{bbm}

\usepackage{mathtools}
\MakeOuterQuote{"}

\newcommand{\amp}{\mathcal{M}}

\newcommand{\beq}{\begin{eqnarray}}
\newcommand{\eeq}{\end{eqnarray}}
\newcommand{\ba}{\begin{eqnarray}}
\newcommand{\ea}{\end{eqnarray}}
\newcommand{\be}{\begin{equation}}
\newcommand{\ee}{\end{equation}}
\newcommand{\bpmatrix}{\begin{pmatrix}}
\newcommand{\epmatrix}{\end{pmatrix}}

\newcommand{\comment}[1]{\ignorespaces}

\newcommand{\s}{\newline \vspace*{-3.5mm}}

\begin{document}

\title{
	\vspace*{-3cm}
	\phantom{h} \hfill\mbox{\small KA-TP-07-2021}
	\vspace*{0.7cm}
\\[-1.1cm]
	\vspace{15mm}   
	\textbf{One-loop Corrections to the Higgs Boson Invisible
          Decay in the Dark Doublet Phase of the N2HDM \\[4mm]}} 
\date{}
\author{
Duarte Azevedo$^{1\,}$\footnote{E-mail: \texttt{drpazevedo@fc.ul.pt}},
Pedro Gabriel$^{1\,}$\footnote{E-mail: \texttt{pedrogabriel347@hotmail.com}},\\
Margarete M\"{u}hlleitner$^{2\,}$\footnote{E-mail:
	\texttt{margarete.muehlleitner@kit.edu}},
Kodai Sakurai$^{2,3\,}$\footnote{Address after April 2021, Department
  of Physics, Tohoku University, Sendai, Miyagi 980-8578, Japan; 
  E-mail: \texttt{kodai.sakurai.e3@tohoku.ac.jp}}, 
Rui Santos$^{1,4\,}$\footnote{E-mail:
  \texttt{rasantos@fc.ul.pt}}
\\[9mm]
{\small\it
$^1$Centro de F\'{\i}sica Te\'{o}rica e Computacional,
    Faculdade de Ci\^{e}ncias,} \\
{\small \it    Universidade de Lisboa, Campo Grande, Edif\'{\i}cio C8
  1749-016 Lisboa, Portugal} \\[3mm]
{\small\it
$^2$Institute for Theoretical Physics, Karlsruhe Institute of Technology,} \\
{\small\it Wolfgang-Gaede-Str. 1, 76131 Karlsruhe, Germany.}\\[3mm]
{\small\it$^3$Institute for Astroparticle Physics, Karlsruhe Institute of Technology,}\\
{\small\it 76344 Karlsruhe, Germany.}\\[3mm]
{\small\it
$^4$ISEL -
 Instituto Superior de Engenharia de Lisboa,} \\
{\small \it   Instituto Polit\'ecnico de Lisboa
 1959-007 Lisboa, Portugal} \\[3mm]
}
\maketitle

%
%

\begin{abstract}
The Higgs invisible decay width may soon become a powerful tool to
probe extensions of the Standard Model with dark matter  
candidates at the Large Hadron Collider. In this work, we calculate the
next-to-leading order (NLO) electroweak corrections to the 125 GeV Higgs 
decay width into two dark matter particles. The model is the
next-to-minimal 2-Higgs-doublet model (N2HDM) in
the dark doublet phase, that is, only one doublet and 
the singlet acquire vacuum expectation values.  We show that the
present measurement of the Higgs invisible branching  
ratio, BR$(H \to$ invisible $< 0.11$), does not lead to constraints
on the parameter space of the model at leading order.  
This is due to the very precise measurements of the Higgs couplings
but could change in the near future.  
Furthermore, if NLO corrections are required not to be
unphysically large, no limits on the parameter space can be extracted from the NLO results.
\end{abstract}

\newpage

\thispagestyle{empty}
\vfill
\newpage

\section{Introduction}
\label{sec:intro}
Ever since the Higgs boson was discovered at the Large Hadron Collider
(LHC) by the ATLAS~\cite{Aad:2012tfa} and
CMS~\cite{Chatrchyan:2012ufa} collaborations, the measurement of the
Higgs couplings to the 
remaining Standard Model (SM) particles became a powerful tool in
constraining the parameter space of extensions of the SM. Another
important ingredient when building extensions of the SM with dark
matter (DM) candidates is the measurement of the invisible Higgs branching
ratio. Very recently, a new result by the ATLAS collaboration combining 139 fb$^{-1}$ of data at $\sqrt s = 13$ TeV
with the results obtained at $\sqrt s = 7$ and $8$ TeV was published. The observed
upper limit on the SM-like Higgs ($H_{\text{SM}}$) to invisibles branching ratio (BR)
is 0.11~\cite{ATLAS:2020kdi}, which is an improvement from
the previous result with an invisible BR above 0.2. The results of
the Higgs coupling measurements together with those of the invisible
Higgs decay are our best tools at colliders to constrain extensions of
the scalar sector of the SM with DM candidates. \s

We will focus on a specific phase of the next-to-minimal 2-Higgs-doublet model
(N2HDM) with a scalar sector consisting of two complex doublets and
one real singlet. Only one of the doublets 
and the singlet acquire vacuum expectation values (VEVs) and we end up
with two possible DM candidates. This particular phase of the N2HDM is
known as the dark doublet phase (DDP). The different phases of the N2HDM are
described in detail in~\cite{Engeln:2020fld}. \s 

Our analysis will be performed by first imposing the most relevant
theoretical and experimental constraints on the model. We then
calculate the next-to-leading (NLO) electroweak corrections 
to the invisible decay of the SM-like Higgs boson, that is, the Higgs
decaying into two DM candidates. The results will be presented for all
allowed parameter space points, which will enable 
us to understand if NLO corrections can help to constrain the
parameter space of the N2HDM. The NLO BR of Higgs to invisibles could
be larger than  the experimentally measured value 
for some regions of the parameter space. \s

As shown in a recent work~\cite{Engeln:2020fld}, the constraints
coming from the Higgs couplings to fermions and gauge bosons are
enough to indirectly constrain the BR of the Higgs decay into 
invisibles to be below 0.1 in the N2HDM (DDP phase). So until recently,
the Higgs BR to invisible was not a meaningful experimental result to
constrain the parameter space. However, 
the new measurement by ATLAS, reaching now 0.11, is exactly at the
frontier between the indirect bound coming from Higgs couplings and
the direct one coming from the invisibles Higgs BR. 
So it is extremely timely to calculate the electroweak (EW) NLO corrections to the
Higgs invisible BR. \s

The outline of the paper is as follows. In section \ref{sec:model}, we
will introduce the DDP phase of the N2HDM together with our
notation. Section~\ref{sec:ren} 
is dedicated to the description of the different renormalization
schemes used in this work. 
In section~\ref{sec:inv} we discuss the expressions of the Higgs
invisible decay at leading order (LO) and at NLO. In section \ref{sec:num}, the results
are presented and discussed. Our conclusions are collected in
section~\ref{sec:conc}. There are two appendices where we discuss
details of the renormalization procedure.

\section{Model}
\label{sec:model}
We start this section by describing the dark doublet phase of the
N2HDM~\cite{Chen:2013jvg,Drozd:2014yla,Jiang:2016jea,Muhlleitner:2016mzt,Engeln:2020fld}.  
The Higgs sector of  the N2HDM is composed of two $SU(2)_{L}$
doublet fields with hypercharge Y=1, $\Phi_{i}$~($i=1,2$), and the real
$SU(2)_{L}$ singlet field $\Phi_{S}$ with hypercharge Y=0.  
Two discrete $\mathbb Z_{2}$ symmetries, $\mathbb Z^{(1)}_{2}$ and $\mathbb
Z^{(2)}_{2}$, are imposed on the model.  
Under these symmetries, each scalar field is transformed as
\begin{align}
\mathbb{Z}^{(1)}_{2}:&\quad \Phi_{1}\rightarrow \Phi_{1},\quad \Phi_{2}\rightarrow -\Phi_{2},\quad \Phi_{S}\rightarrow \Phi_{S}\,,\label{eq:Z2_1} \\
\mathbb{Z}^{(2)}_{2}:&\quad \Phi_{1}\rightarrow \Phi_{1},\quad \Phi_{2}\rightarrow \Phi_{2},\quad \Phi_{S}\rightarrow -\Phi_{S}\label{eq:Z2_2} \,.
\end{align}
We require the $\mathbb Z_{2}$ symmetries to be exact, meaning that
no soft breaking terms are introduced, and therefore the Higgs
potential of the N2HDM is given
by~\cite{Chen:2013jvg,Drozd:2014yla,Jiang:2016jea,Muhlleitner:2016mzt,Engeln:2020fld} 
\begin{align}
\label{eq:scalpot}
V_{} =&\enspace m_{11}^{2} \Phi_{1}^{\dagger} \Phi_{1} + m_{22}^{2} \Phi_{2}^{\dagger} \Phi_{2}
+ \dfrac{\lambda_{1}}{2} \left(\Phi_{1}^{\dagger} \Phi_{1}\right)^{2}
+ \dfrac{\lambda_{2}}{2} \left(\Phi_{2}^{\dagger} \Phi_{2}\right)^{2}\notag\\
&+\enspace \lambda_{3} \Phi_{1}^{\dagger} \Phi_{1} \Phi_{2}^{\dagger} \Phi_{2}
+ \lambda_{4} \Phi_{1}^{\dagger} \Phi_{2} \Phi_{2}^{\dagger} \Phi_{1}
+ \dfrac{\lambda_{5}}{2} \left[\left(\Phi_{1}^{\dagger} \Phi_{2}\right)^{2} + \text{h.c.}\right]\\ 
&+\enspace \dfrac{1}{2} m_{s}^{2}\Phi_{S}^{2} + \dfrac{\lambda_{6}}{8} \Phi_{S}^{4} + \dfrac{\lambda_{7}}{2} \Phi_{1}^{\dagger} \Phi_{1} \Phi_{S}^{2} + \dfrac{\lambda_{8}}{2} \Phi_{2}^{\dagger} \Phi_{2} \Phi_{S}^{2}\,, \notag
\end{align}
where all parameters can be set real by rephasing $\Phi_{1}$ or
$\Phi_{2}$. 
In the N2HDM, there are four different minima, which break the $SU(2)\times U(1)_{Y}$ symmetry into $U_{\rm EM}(1)$, depending on the vacuum expectation values  for the doublet fields and the singlet field, i.e. $\left<\Phi_{1}\right>,\ \left<\Phi_{2}\right>$ and $\left<\Phi_{S}\right>$, respectively.  
The possible patterns are 
\begin{align}
\mbox{broken phase (BP):}& 
\left<\Phi_{1}\right>\neq0\,, \left<\Phi_{2}\right>\neq0\,, \left<\Phi_{S}\right>\neq0\,,  \\
\mbox{dark doublet phase (DDP):}&
\left<\Phi_{1}\right>\neq0\,, \left<\Phi_{2}\right>=0\,, \left<\Phi_{S}\right>\neq0 
 \,,  \\
\mbox{dark singlet phase (DSP):}&
\left<\Phi_{1}\right>\neq0\,, \left<\Phi_{2}\right>\neq0\,, \left<\Phi_{S}\right>=0\,,  \\
\mbox{full dark phase (FDP):}&
\left<\Phi_{1}\right>\neq0\,, \left<\Phi_{2}\right>=0\,, \left<\Phi_{S}\right>=0  
\,.
\end{align}
In this study, we focus on the DDP, where $\mathbb Z^{(1)}_{2}$ remains unbroken while $\mathbb Z^{(2)}_{2}$ is spontaneously broken. 
Hence, this phase corresponds to an extension of the inert doublet
model (IDM)~\cite{Deshpande:1977rw} by the additional singlet field. 
The other phases are discussed in
Refs.~\cite{Chen:2013jvg,Drozd:2014yla,Jiang:2016jea,Muhlleitner:2016mzt,Engeln:2020fld}. \s

In the DDP, the components of the Higgs fields can be parameterized as
 \begin{align}
\Phi_1 = \begin{pmatrix} G^+ \\
\dfrac{1}{\sqrt{2}}\left(v + \rho_{1} + i\, G^{0}\right)
\end{pmatrix},\quad
\Phi_2 = \begin{pmatrix} H^+_{D} \\
\dfrac{1}{\sqrt{2}}\left(H_{D} + i\,A_{D}\right)
\end{pmatrix}, \quad
\Phi_S = v_{s} + \rho_{s}\,,
\end{align}
where $v=246$ GeV is the electroweak VEV and $v_{S}$ is the VEV of the singlet field. 
The doublet field $\Phi_1$ corresponds the SM Higgs doublet, which
contains the Nambu-Goldstone bosons $G^+$ and $G^{0}$. 
Due to the unbroken $\mathbb Z^{(1)}_{2}$ symmetry, the four dark
scalars, $H_{D}, A_{D}$ and $H^\pm_{D}$ do not mix, i.e., they
are physical states.  
The lightest neutral dark scalar, which can be either $H_{D}$ or
$A_{D}$, is the DM candidate.  
On the other hand, the two CP-even Higgs fields $\rho_{1}$ and $\rho_{S}$
mix with each other.  
Together with the CP-even dark scalar $H_{D}$, the mass eigenstates
for the CP-even Higgs bosons can be expressed through a rotation
matrix with the mixing angle $\alpha$ as, 
\begin{align}
\label{eq:NH}
\left(\begin{array}{c}H_1 \\H_2 \\H_D\end{array}\right)
=\left(\begin{array}{ccc}c_\alpha & 0 & s_\alpha \\-s_\alpha & 0 & c_{\alpha} \\0 & 1 & 0\end{array}\right)
\left(\begin{array}{c}\rho_1 \\ H_{D} \\ \rho_S\end{array}\right) , 
\end{align}
where by convention, we take $m_{H_{1}}<m_{H_{2}}$, and where we have
introduced the short-hand notations $c_\alpha \equiv \cos\alpha$ and
$s_\alpha \equiv  \sin\alpha$. 
Either $H_{1}$ or $H_{2}$ can be identified as the SM-like Higgs boson
($H_{\text{SM}}$) with a mass of 125~GeV.  
For later convenience, we define the rotation matrix as $R$, so that
Eq. \eqref{eq:NH} can be rewritten by
$H_{i}=R_{ij}\rho_{j}$ ($i,j=1,2,3$), defining $\rho_2 = H_D$ and $\rho_3 = \rho_S$.  \s

The masses of the physical states can be written as
\begin{align}
\label{eq:ID_mH1sq}
m^2_{H_{1}} 
&=  v^2 \cos^2\alpha\, \lambda_1 + v_s^2 \sin^2\alpha\, \lambda_6 + 2v v_s \sin\alpha\,\cos\alpha\, \lambda_7\,,\\
\label{eq:ID_mH2sq}
m^2_{H_{2}} 
 &= v^2 \sin^2\alpha\,\lambda_1 + v_s^2 \cos^2\alpha\,\lambda_6 - 2 v v_s \sin\alpha\,\cos\alpha\, \lambda_7\,,\\
\label{eq:ID_mHDsq}
m^2_{H_{D}} &= \dfrac{1}{2} (2 m_{22}^2 + v^2 (\lambda_3 + \lambda_4 + \lambda_5) + v_s^2 \lambda_8)\,,\\
\label{eq:ID_mADsq}
m^2_{A_{D}}
&= \dfrac{1}{2} (2 m_{22}^2 + v^2 (\lambda_3 + \lambda_4 - \lambda_5) + v_s^2 \lambda_8)\,,\\
\label{eq:ID_mHcDsq}
m^2_{H_{D}^{\pm}}
&= \dfrac{1}{2} (2 m_{22}^2 + v^2 \lambda_3 + v_s^2 \lambda_8)\,.
\end{align}
Using these mass formulae together with the mixing angle $\alpha$ and the stationary conditions for $\Phi_{1}$ and $\Phi_{S}$, i.e., 
\begin{eqnarray}
\left<\partial V/\partial\Phi_{1}\right>\equiv T_{\Phi}/v =0 \quad
  \mbox{and} \quad
  \left<\partial V/\partial\Phi_{S}\right>\equiv T_{S}/v_{S} =0 \;,
\end{eqnarray} 
the original parameters of the potential can be replaced  by a new set to be used as input. Together with the electroweak and singlet VEVs, we choose as our input the
following 13 parameters, 
\begin{equation}
\label{eq:inpparaDDP}
   v\,,\enspace v_s\,,\enspace
  m_{H_{1}}\,,\enspace m_{H_{2}}\,,\enspace m_{H_{D}}\,,\enspace
  m_{A_D}\,,\enspace m_{H_D^{\pm}}\,, \enspace
   \alpha\,,\enspace m_{22}^2\,,\enspace T_{\Phi} \,,\enspace T_{S} \,,\enspace \lambda_2\,,\enspace \lambda_8\, .
  \end{equation}

We assign  $\mathbb Z^{(1)}_{2}$-even and $\mathbb Z^{(2)}_{2}$-even
parity to the remaining SM fields and consequently only the Higgs
doublet $\Phi_{1}$ has Yukawa interactions. This in turn means that  
the Yukawa couplings are just the SM ones and that the dark scalars do not
couple to the SM fermions. 
On the other hand, due to the kinetic term for $\Phi_{2}$, \textit{dark
scalar-dark scalar-gauge boson} type of vertices are allowed while the
\textit{dark scalar-gauge boson-gauge boson} type of vertices are forbidden by the
$\mathbb Z^{(1)}_{2}$ symmetry.  
The Feynman rules for the vertices including two dark scalars and a
gauge boson are written in Ref~\cite{Engeln:2020fld}. In particular,
the trilinear scalar couplings that are relevant for the calculation of the
invisible decay  of the Higgs boson are given by ($i=1,2$)
 \begin{align}
\label{eq:lam1DD}
\lambda_{H_{i}H_{D}H_{D}}=-\frac{1}{v}\big[
2(m_{H_{D}}^{2}-m_{22}^{2})R_{i1}+\lambda_{8} v_{S}(vR_{i3}-v_{S}R_{i1})
\big] \;.
\end{align}
We note that in the case of $\cos\alpha=1$ and $\lambda_{8}=0$ the expression of $\lambda_{H_{i}H_{D}H_{D}}$ is exactly the IDM one. 
The other trilinear scalar couplings are also given in Ref.~\cite{Engeln:2020fld}. 

\section{Renormalization }
\label{sec:ren}

In this section, we discuss the renormalization scheme used in the
calculation of the one-loop corrections to the 
Higgs boson $H_i$ $(i=1,2)$ decay into a pair of DM particles,  which we assume to be $H_D$, unless otherwise stated, $H_i \to H_D
  H_D$. Thus, this section focuses on the renormalization of the scalar and gauge sectors. 
The renormalization of the fermion sector as well as any treatment of
infrared divergence is not necessary for this particular process. \s

We perform the renormalization of the Higgs sector in the DDP of the
N2HDM according to the procedure presented in
Ref.~\cite{Krause:2016oke} for the 2HDM and in
Ref.~\cite{Krause:2017mal} for the broken 
phase of the N2HDM. 
Although most of the parameters  in the Higgs sector of the DDP are common with those of the broken phase, we describe the renormalization of all parameters in order to make the paper self-contained. 

\subsection{Gauge sector}
 The renormalization of all parameters and fields in the gauge sector
 is done using the on-shell (OS) scheme following Ref.~\cite{Denner:1991kt}. 
 As the three independent parameters in this sector, we choose the
 masses of weak gauge bosons and the electric charge, i.e., $m_{W}$,
 $m_{Z}$, and $e$, respectively.  
 These parameters are shifted as
 \begin{align}
 m_{V}^{2}&\to  m_{V}^{2} +\delta m_{V}^{2} \quad (V=W,Z) \;, \\
 e&\to (1+\delta Z_{e})e \;. 
 \end{align}
 Moreover, the bare fields for the gauge bosons in the mass basis
 are replaced by the renormalized ones as 
 \begin{align}
 W^{\pm}&\to  (1+\frac{1}{2}\delta Z_{WW})W^{\pm}\,, \notag \\
 \begin{pmatrix}
 Z\\
\gamma
 \end{pmatrix}
 &\to
 \begin{pmatrix}
 1+\frac{1}{2}\delta Z_{ZZ}&\frac{1}{2}\delta Z_{Z\gamma} \\
\frac{1}{2}\delta Z_{\gamma Z}  &1+\frac{1}{2}\delta Z_{\gamma\gamma}
 \end{pmatrix}
 \begin{pmatrix}
 Z\\
\gamma
 \end{pmatrix}\,.
  \end{align}
%
The OS conditions for these gauge fields are defined as
\beq
\delta m_W^2 = \mbox{Re} \Sigma^{\text{tad},T}_{WW} (m_W^2) \quad
\mbox{and} \quad
\delta m_Z^2 = \mbox{Re} \Sigma^{\text{tad},T}_{ZZ} (m_Z^2) \;,
\eeq
\beq
\delta Z_{WW} &=& - \mbox{Re} \left.\frac{\partial \Sigma^T_{WW} 
  (p^2)}{\partial p^2}\right|_{p^2=m_W^2}\,, \label{eq:sigmaww} \\
\left( \begin{array}{cc} \delta Z_{ZZ} & \delta Z_{Z\gamma} \\ \delta Z_{\gamma Z} &
  \delta Z_{\gamma\gamma} \end{array} \right) &=&
\left( \begin{array}{cc} - \mbox{Re} \left.\frac{\partial \Sigma^T_{ZZ}
  (p^2)}{\partial p^2}\right|_{p^2=m_Z^2} & 
2 \mbox{Re}\frac{\Sigma^T_{Z\gamma} (0)}{m_Z^2} \\
 -2 \mbox{Re} \frac{\Sigma^T_{Z\gamma} (m_Z^2)}{m_Z^2} & 
- \left.\frac{\partial \Sigma^T_{\gamma\gamma} (p^2)}{\partial
    p^2}\right|_{p^2=0} \end{array}\right) \; \label{eq:sigmagamz},
\eeq
where $\Sigma^{\text{tad},T}_{WW}$ and $\Sigma^{\text{tad},T}_{ZZ}$ denote the transverse part of the self-energies of the $W$ and $Z$ bosons. These contain
the tadpole contributions due to our renormalization scheme choice. No "tad" superscript means that there is no contribution from the tadpole diagrams. The different 
tadpole schemes will be described below. 
 %
The counterterm for the electric charge is determined from
$\gamma e \bar{e}$ in the Thomson limit and can be expressed as a function of the self-energies as
  \beq
\delta Z_e^{\alpha(0)} &=& \frac{1}{2} \left.\frac{\partial
  \Sigma^T_{\gamma\gamma} (k^2)}{\partial k^2}\right|_{k^2=0} +
\frac{s_W}{c_W} \frac{\Sigma_{\gamma Z}^T (0)}{m_Z^2} \;.
\label{eq:deltaze0}
\eeq
This counterterm contains large logarithmic corrections arising from
the small fermion masses, $\log m^{2}_{f}~(f\neq t)$. We use the "$G_\mu$ scheme"
  \cite{Bredenstein:2006rh}  
  in order to improve the perturbative behaviour. In this scheme, 
a large universal part of the ${\cal O}(\alpha)$ corrections is
absorbed in the leading order decay width by deriving the electromagnetic
coupling constant $\alpha=e^2/(4\pi)$ from the Fermi constant, $G_\mu$, as
\beq
\alpha_{G_\mu} = \frac{\sqrt{2}G_\mu m_W^2}{\pi} \left( 1-
  \frac{m_W^2}{m_Z^2} \right) \;.
\eeq
This allows us to take into account the running of the electromagnetic
coupling constant $\alpha (Q^2)$, from $Q^2=0$ to the electroweak
scale. In order to avoid double counting, the corrections that are
absorbed in the LO decay width by using $\alpha_{G_\mu}$ have to
be subtracted from the explicit ${\cal O}(\alpha)$ corrections. This
is achieved by subtracting the weak corrections to the muon decay, $\Delta r$
\cite{Denner:1991kt,Sirlin:1980nh}, from the corrections in the
$\alpha(0)$ scheme. Hence, we redefine the charge renormalization constant as
\beq
\delta Z_e |_{G_\mu} = \delta Z_e |_{\alpha(0)} - \frac{1}{2} (\Delta
r)_{\text{1-loop}} \;, \label{eq:redef}
\eeq
where $(\Delta r)_{\text{1-loop}}$ is the one-loop expression for
$\Delta r$ given by \cite{Denner:1991kt}
\beq
(\Delta r)_{\text{1-loop}} &=& \left. \frac{\partial
  \Sigma^T_{\gamma\gamma} (k^2)}{\partial k^2}\right|_{k^2=0} -
\frac{c_W^2}{s_W^2} \left( \frac{\Sigma_{ZZ}^T (m_Z^2)}{m_Z^2}-
\frac{\Sigma_{WW}^T (m_W^2)}{m_W^2} \right) + \frac{\Sigma_{WW}^T (0)-
\Sigma_{WW}^T (m_W^2)}{m_W^2} \nonumber \\
&& - 2 \frac{c_W}{s_W} \frac{\Sigma_{\gamma Z}^T (0)}{m_Z^2}
+ \frac{\alpha}{4\pi s_W^2} \left( 6 + \frac{7-4s_W^2}{2s_W^2} \log
  c_W^2 \right) \;.
\eeq
Note that through the redefinition Eq.~(\ref{eq:redef}), the first term
of $\delta Z_e^{\alpha(0)}$ in Eq.~(\ref{eq:deltaze0}), which contains
the large logarithmic corrections from the light fermion loops,
cancels against the corresponding term in $(\Delta r)_{\text{1-loop}}$.
%
The counterterms for the other EW parameters can be expressed in terms of those presented above.  
For example, the $SU(2)_{L}$ gauge coupling, $g$, is replaced by the
tree level relation $g=em_{Z}/\sqrt{m_{Z}^{2}- m_{W}^{2}}$. Thus, the counterterm is given by
\begin{align}
\frac{\delta g}{g}=\delta {Z_{e}}-\frac{1}{2(1-m^{2}_{Z}/m^{2}_{W})}
\left(\frac{\delta m_{W}^{2}}{m_{W}^{2}} - \frac{\delta m_{Z}^{2}}{m_{Z}^{2}} \right)\,.
\end{align}
\subsection{Higgs sector}
In the Higgs sector, we have a total of 13 free parameters, given in
Eq.~\eqref{eq:inpparaDDP}, considering the two tadpoles $T_{\Phi}$ and $T_{S}$. 
We have to renormalize the scalar fields in the mass basis, $H_{1}$, $H_{2}$, $H_{D}$, $A_{D}$ and $H^{\pm}_{D}$. 
%
The counterterms are introduced via the shift of the input parameters,
i.e, the masses of the scalar bosons, the mixing angle $\alpha$ of
the CP-even Higgs bosons and the remaining original potential
parameters that appear in the vertices of the processes under study,
$\lambda_8$ and $m_{22}^2$, 
\begin{align}
&m_{\Phi}^{2}\to m_{\Phi}^{2}+\delta m_{\Phi}^{2}\,,\quad
\alpha\to \alpha+ \delta \alpha\,,  \notag\\
&m_{22}^{2}\to m_{22}^{2}+\delta  m_{22}^{2} \,, \quad	
\lambda_{8}\to \lambda_{8}+\delta \lambda_{8}\,,
\end{align}
where $\Phi$ denotes $H_{1},\ H_{2},\ H_{D},\ A_{D},$ and
$H^{\pm}_{D}$. There is no need to
renormalize $\lambda_2$ for this particular process. Apart from the tadpoles, the remaining two parameters are the
VEVs. The electroweak VEV 
$v$ is fixed by the $W$ mass and the renormalization of $v_S$ will be
discussed later. \s

The tadpole renormalization can be performed in different ways and we
will discuss two approaches. These are designated by Standard Tadpole Scheme (STS) 
and Alternative Tadpole Scheme (ATS). The latter was originally proposed
by Fleischer and Jegerlehner, in Ref.~\cite{Fleischer:1980ub}, for the
SM. The ATS was also discussed in
detail for the CP-conserving 2HDM in Ref.~\cite{Krause:2016oke} and
for  
the broken phase of the N2HDM in Ref.~\cite{Krause:2017mal}. We will
just briefly review the two schemes for completeness. \s

In the STS, the tree level tadpoles are replaced by 
\begin{align}
T_{X} \to T_{X}+\delta T_{X} \quad(X=\Phi, S)\,,
\end{align}
and are chosen as the renormalization parameters.
On the other hand, in the ATS, the VEVs are the renormalization parameters and are shifted as 
\begin{align}\label{eq:siftvev}
v\to v +\delta v\,,\quad v_{S}\to v_{S}+\delta v_{S}\,.
\end{align}
We use the ATS, which will now be explained in more detail. The reason to use this scheme is that, as shown by Fleischer and Jegerlehner,
all renormalized parameters are gauge independent except for the wave function renormalization constants (or any parameter that depends on the wave function 
renormalizaton constants as is the case of the angle $\alpha$ in
particular schemes). \s


Before moving to the discussion of the tadpole renormalization, we define the wave function renormalization constants of the scalar fields. 
The bare fields are replaced by the renormalized ones through 
\begin{align}
\label{eq:WFRCH}
&\left(\begin{array}{c}H_1 \\H_2 \\H_D\end{array}\right)
\to 
\left(\begin{array}{ccc}1+\frac{1}{2}\delta Z_{H_{1}H_{1}} & \frac{1}{2}\delta Z_{H_{1} H_{2}} & 0 \\ \frac{1}{2}\delta Z_{H_{2} H_{1}} & 1+\frac{1}{2}\delta Z_{H_{2}H_{2}} & 0 \\0 & 0 & 1+ \frac{1}{2}\delta Z_{H_{D}H_{D}}\end{array}\right)
\left(\begin{array}{c}H_1 \\H_2 \\H_D\end{array}\right)\,, \\ 
&A_{D}\to A_{D}(1 +\frac{1}{2}\delta Z_{A_{D}})\,, \quad H^{\pm}_{D}\to H^{\pm}_{D}(1 +\frac{1}{2}\delta Z_{H^{\pm}_{D}})\,.
\end{align}
Note that, for Eq.~\eqref{eq:WFRCH}, the exact $\mathbb  Z^{(1)}_{2}$ symmetry ensures that the $(3$-$k)$ and $(k$-$3)$
components ($k=1,2$) are zero.  

\subsubsection{Tadpoles }

In the ATS, the renormalized VEVs, which correspond to minima of the Higgs potential at loop-level, are regarded as the tree-level VEVs, namely, one imposes
\begin{align}\label{eq:renoVEVFJ}
 v^{\rm bare}&=v^{\rm ren}+\delta v^{} \overset{\rm FJ}{=} v^{\rm tree}+\delta v^{}\,, \notag \\
 v^{\rm bare}_{S}&= v^{\rm ren}_{S}+\delta v^{}_{S} \overset{\rm FJ}{=} v^{\rm tree}_{S}+\delta v^{}_{S}\,. \quad
\end{align}
Using the tadpole conditions, one can derive expressions for $\delta v$ and $\delta v_{S}$,
\begin{align}\label{eq:renotad}
T_{i}^{\rm bare}= T_{i}^{\rm tree}+f(\delta v_{},\delta v_{S})= T_{i}^{\rm loop}\quad (i=\Phi,S) \,,
\end{align}
where the first term, $T_{i}^{\rm tree}$, is zero from the stationary
condition at the tree-level and the second term, $f(\delta v_{},\delta
v_{S})$, denotes contributions from $\delta v$ and $\delta {v_{S}}$,
which can be extracted by inserting Eq.~\eqref{eq:siftvev} into the
tree-level tadpole conditions.  
From Eq.~\eqref{eq:renotad} one obtains the following expressions for the VEV counterterms
\begin{equation}
{\cal R}(\alpha)
 \left( 
\begin{array}{c} \delta v \\   \delta v_S  
\end{array} 
\right) =\,  \left( 
\begin{array}{c} \frac{T_{H_1}^{\text{loop}}}{m^2_{H_1}} \\
  \frac{T_{H_2}^{\text{loop}}}{m^2_{H_2}} 
\end{array} 
\right) 
\label{eq:VEVCT}\, ,
\end{equation}
where ${\cal R}(\alpha)$ denotes the $2\times2$ non-diagonal part of $R$ (see  Eq.~(\ref{eq:NH})),
\beq
{\cal R}_{11}={\cal R}_{22}= \cos \alpha \;, \quad
{\cal R}_{12}=-{\cal R}_{21}= \sin \alpha \;,
\eeq
and the one-loop tadpoles  in the mass basis are given by $T_{H_{i}}^{\rm loop}=({\cal R}(\alpha))_{ij}T_{j}^{\rm loop}$ ($i,j=1,2$)
  with $T_1=T_\Phi$ and $T_2= T_S$.
The left-hand side corresponds to the VEV counterterms $\delta
v_{H_{1}}$ and $\delta v_{H_{2}}$ in the mass basis and the
right-hand side coincides with the tadpole diagrams multiplied by the
propagator  
for the Higgs bosons at zero momentum transfer, i.e. $T_{H_{i}}^{\rm
  loop}/m_{H_{i}}^{2}=(iT_{H_{i}}^{\rm
  loop})(-i/m_{H_{i}}^{2})$. Therefore, Eq.~\eqref{eq:VEVCT} shows
that $\delta v_{H_{i}}$ can be regarded as the connected tadpole
diagrams for $H_{i}$. 
Once the counterterms for the VEVs are fixed, the shift is performed in all VEV terms in the Lagrangian.
Hence in the ATS, one needs to insert tadpole diagrams in all amplitudes for which the original vertices contain one of the VEVs in addition to the usual one-particle irreducible diagrams.
This general consequence is shown by focusing on specific amplitudes
in Ref.~\cite{Krause:2016oke} for the 2HDM and in Ref.~\cite{Krause:2017mal} for the N2HDM. \s

Another important feature of the ATS is that, because the renormalized VEV is identified with the tree level VEV, the VEVs still have to be renormalized. 
For the EW VEV, the renormalized parameter is given by
\begin{align}\label{eq:renoVEV}
v^{\rm ren}=v^{\rm tree}=2\left.\frac{m_{W}}{g}\right|^{\rm tree}\,,
\end{align}
and the tree-level parameters $g$ and $m_{W}$ are shifted as
\begin{align}\label{eq:DelveEWv}
2\left.\frac{m_{W}}{g}\right|^{\rm tree} &\to 2\left.\frac{m_{W}}{g}\right|^{\rm ren}+\frac{2m_{W}}{g}\left(\frac{\delta m_{W}^{2} }{2m_{W}^{2}}-\frac{\delta g}{g}\right) \notag \\
&\equiv 2\left.\frac{m_{W}}{g}\right|^{\rm ren} +\Delta v\,.
\end{align}
We have defined the shift of the tree level parameters related to the EW VEV as $\Delta v$, which has no relation with $\delta v$. 
The same discussion holds for the singlet VEV $v_{S}$. Once $v_{S}$ is related with some measurable quantity, a similar relation with Eq.\eqref{eq:renoVEV} 
must exist, even if a physical process has to be used, and then $\Delta v_{S}$ has to be introduced. 
\subsubsection{Mass and Wave Function Renormalization  }

The counterterms for the masses and the wave function renormalization
constants (WFRCs)  are determined by imposing the on-shell conditions
 for each scalar field.  
This yields the mass counterterms
\begin{align}
\delta m_{\Phi}^{2} &= \Sigma^{\rm tad}_{\Phi \Phi}(m_{\Phi}^{2})\,, 
\end{align}
and the WFRCs
\begin{align}
\delta Z_{\Phi \Phi} &= - \frac{\partial \Sigma^{\rm tad}_{\Phi \Phi}(p^{2}) }{\partial p^{2}}\Big|_{p^{2}=m_{\Phi}^{2}} \,, \\ 
\delta Z_{H_{i}H_{j}} &= 2\frac{\Sigma^{\rm tad}_{H_{i}H_{j}}(m_{H_{j}}^{2}) }{m_{H_{i}}^{2}-m_{H_{j}}^{2}} \quad (i=1,2, j\neq i) \,,
\end{align}
where, to reiterate, $\Sigma^{\rm tad}$
stands for the self-energy containing one particle irreducible (1PI) diagrams and tadpole contributions. 

\subsubsection{Mixing angle  $\alpha$}

The renormalization of the mixing angle, $\alpha$, requires special
treatment since the gauge dependence of $\delta \alpha$ 
could result in a gauge-dependent physical process~\cite{Yamada:2001px, Espinosa:2002cd, Sperling:2013eva,Bojarski:2015kra, Krause:2016xku, Denner:2016etu, Altenkamp:2017ldc, Kanemura:2017wtm, Fox:2017hbw, Denner:2018opp, Dudenas:2020ggt}. 
 A gauge-independent amplitude can be
  obtained by starting with a gauge-independent definition of $\delta \alpha$.
One possible solution to avoid this gauge dependence is to apply the
ATS for the renormalization of the tadpoles and to make use of the
pinch technique~\cite{Cornwall:1989gv, Papavassiliou:1994pr},  
while keeping the on-shell renormalization for the mixing angle.  
This is the procedure that we adopt throughout this paper. 
\s 

The expression for $\delta \alpha$ in the OS scheme can be derived by
relating quantities in the gauge basis to the corresponding ones in
the physical (mass) basis. This procedure is described in detail
in~\cite{Pilaftsis:1997dr, Kanemura:2004mg, Krause:2016oke}. Following~\cite{Krause:2016oke}  leads to the following expression for $\delta \alpha$ after
adding the pinched terms,
\begin{align}
\delta \alpha &= \frac{1}{4}(\delta Z_{H_{1}H_{2}}-\delta Z_{H_{2}H_{1}})\notag \\
&+\frac{1}{2}\frac{1}{m_{H_{1}}^{2}-m_{H_{2}}^{2} }
\left[\Sigma_{H_1H_2}(m_{H_2}^{2})+\Sigma_{H_{1}H_{2}}(m_{H_{1}}^{2}) \right]\,, 
\end{align}
where $\Sigma_{H_1H_2}$ stands for pinched
contributions to the $H_{1}-H_{2}$ mixing self-energy,  
which remove the gauge-dependent part coming from the first two terms.  
They can be extracted from the expressions obtained for the broken
phase of the N2HDM (see~\cite{Krause:2017mal}) as 
\begin{align}
\Sigma_{H_{1}H_{2}}(q^{2})=-\frac{g^{2}}{32\pi^{2}c_{W}^{2}}c_{\alpha}s_{\alpha}\left(q^{2}-\frac{m_{H_{1}}^{2}+m_{H_{2}}^{2} }{2}\right) \notag \\
\times\left\{B_{0}(q^{2};m_{Z}^{2},m_{Z}^{2})+2c_{W}^{2}
B_{0}(q^{2};m_{W}^{2},m_{W}^{2}) \right\}\,. 
\end{align}
The expression was obtained with the replacement $(\beta-\alpha_{1},\alpha_{2},\alpha_{3})\to(\alpha,0,0)$ and the fact that the $(A_D, Z)$ and $(H_D^{\pm}, W)$  loop contributions do not appear in our calculation due to the existence of an exact $\mathbb  Z^{(1)}_{2}$ symmetry.  

\noindent 
\subsubsection{ Counterterms for ${\lambda_{8}}$ and $ m_{22}^{2}$ }

The counterterms for the quartic coupling ${\delta \lambda_{8}}$ and
the invariant mass for the dark scalars  ${\delta m_{22}^{2}}$ cannot
be renormalized using OS conditions for the Higgs states. 
Hence we will renormalize these parameters using three different schemes: the $\overline{\rm MS}$ scheme, a process-dependent scheme and one derivation
of the latter that consists of taking the external momenta to be zero
instead of taking them on-mass-shell. \s 
 
In the $\overline{\rm MS}$ scheme, the analytic expressions for the
counterterms can be extracted from the beta functions at one-loop,
yielding 
\begin{align}\label{eq:CTlam8m22MS}
\delta \lambda_{8}=\frac{1}{32\pi^{2}}\beta_{\lambda_{8}}^{(1)} \Delta_{\rm div} \,, \ \ \ 
\delta m_{22}^{2}=\frac{1}{32\pi^{2}}\beta_{m_{22}^{2}}^{(1)}\Delta_{\rm div} \,, 
\end{align}
where $\Delta_{\rm div}$ denotes the UV-divergent part, i.e.,
$\Delta_{\rm div} =1/\epsilon -\gamma_{E}+\log(4\pi)$. 
$\gamma_{E}$ is the Euler-Mascheroni constant and $1/\epsilon$ is the
UV pole in dimensional regularisation.  
The beta functions are given in terms of the original potential
parameters by
\begin{align}
\beta_{\lambda_8}^{(1)} & =  
2 \lambda_4 \lambda_7  + 4 \lambda_3 \lambda_7  + \frac{1}{10} \lambda_8 \Big(30 \lambda_6  + 40 \lambda_8  -45 g_{2}^{2}  + 60 \lambda_2  -9 C^{2}g_{1}^{2} \Big)\,,  \\ 
\beta_{m^2_{22}}^{(1)} & =  
2 \lambda_4 m^2_{11}  + 4 \lambda_3 m^2_{11}  + 6 \lambda_2 m^2_{22}  -\frac{9}{10} C^{2} g_{1}^{2} m^2_{22}  -\frac{9}{2} g_{2}^{2} m^2_{22}  + \lambda_8 m^2_S\,,  
\end{align}
where the Clebsch-Gordan coefficient $C$ is given by $C^{2}=5/3$ and $g_1$ and $g_2$ denote the $U(1)_Y$ and $SU(2)_L$ gauge
  couplings, respectively. These expressions were derived using  \texttt{SARAH-4.14.2}~\cite{Staub:2009bi,Staub:2010jh,Staub:2012pb,Staub:2013tta,Staub:2015kfa}. \s

For the process-dependent scheme, one can fix the counterterms
${\delta \lambda_{8}}$
and  ${\delta m_{22}^{2}}$, required in the one-loop decays $H_i \to
H_D H_D$ $(i=1,2)$, by making use of the Higgs boson decays
into a pair of dark CP-odd scalars. 
We choose as renormalization condition that the decay width  for
$H_{i}\to A_{D} A_{D}$ at NLO coincides with that at LO, namely, 
\begin{align}\label{eq:procreno}
\Gamma_{H_{i}\to A_{D} A_{D}}^{\rm LO}
\overset{!}{=}
 \Gamma_{H_{i}\to A_{D} A_{D}}^{\rm NLO} \quad (i=1,2)\,.
\end{align}
The counterterms ${\delta \lambda_{8}}$ and  ${\delta m_{22}^{2}}$
defined by these conditions contain not only UV-divergent parts but
also finite terms.  
The detailed explanation on how the counterterms are computed is
given in Appendix~\ref{ap:lam8m22}. \s 

The process-dependent scheme takes all particles to be on-shell because it uses a physical process. This means, however, that the
renormalization conditions Eq.~\eqref{eq:procreno} can only be used  
if the decay processes $H_{i}\to A_{D} A_{D}$ are kinematically
allowed. There is a way to circumvent this problem by not taking the
particles on-shell. \s

The renormalization conditions  Eq.~\eqref{eq:procreno} can be written as,
\begin{equation}\label{eq:PDS}
2{\rm Re}\big(\mathcal{M}_{i}^{{\rm tree}*}\mathcal{M}_{i}^{{\rm 1\mathchar`-loop}}\big)\Big|_{p_i^2 = m_i^2,\ p_{A_{D}}^2 = m_{A_{D}}^2} = 0  \qquad \Rightarrow \qquad {\rm Re} \big(\mathcal{M}_{i}^{{\rm 1\mathchar`-loop}}\big)\Big|_{p_i^2 = m_i^2 ,\ p_{A_{D}}^2 = m_{A_{D}}^2} = 0 
\end{equation}
because the tree-level amplitude is just a real constant. If instead we choose to use the same condition but with all external momenta equal to zero, we will not be restricting the parameter space of the model
that can be probed. The third renormalization scheme is therefore defined by  
\begin{equation}\label{eq:newPDS}
\big(\mathcal{M}_{i}^{{\rm 1\mathchar`-loop}}\big)\Big|_{p_i^2 =\ p_{A_{D}}^{2} = 0} = 0 
\end{equation}
while using exactly the same two processes that were used for the
process-dependent scheme. Note that the problem 
in the on-shell case is related to the calculation of $C_0$ loop functions
in forbidden kinematical regions~\cite{tHooft:1978jhc}. 
We will refer to the two
schemes as OS process-dependent and zero external momenta (ZEM) process-dependent in the following.

\subsubsection{Determination of $\Delta v_{S}$ }\label{sec:DvS}
The quantity $\Delta v_{S}$, which is introduced by a similar relation
to the one for the SM in Eq.~\eqref{eq:DelveEWv}, is necessary to get
a UV-finite result for the processes of interest, $H_{i}\to
H_{D}H_{D}$ $(i=1,2)$.  
We note, however, that the renormalization of $ v_{S}$ is only needed
when the parameters ${ \lambda_{8}}$ and  ${m_{22}^{2}}$ are
renormalized via the $\overline{\rm MS}$ scheme conditions. When the process-dependent scheme is used to renormalize ${ \lambda_{8}}$ and  ${m_{22}^{2}}$, the terms with $\Delta v_{S}$ disappear in the renormalized 1-loop amplitude $\mathcal{M}_{H_{i}\to H_{D} H_{D}}^{\rm 1\mathchar`-loop}$, hence in this case $\Delta v_S$ is not necessary.\s

For the $\overline{\rm MS}$  case, our choice is such 
that the remaining UV-divergent part in the  
renormalized amplitude $\mathcal{M} (H_{1}\to H_{D}H_{D})$, which is
not removed by all other counterterms in this vertex, is absorbed  by
$\Delta v_{S}$.  
This results in the following condition
\begin{align}\label{eq:CTDvS}
\Delta v_{S}=-(\delta v_{S})_{\rm div}\,,
\end{align}
as given in Appendix~\ref{ap:DvS}. We checked that by using
Eq.~\eqref{eq:CTDvS} the one-loop amplitude for $H_{2}\to H_{D}H_{D}$ is
also UV-finite. 

\section{The Invisible Higgs Boson Decays at NLO EW}
\label{sec:inv}
In this section, we calculate  the one-loop corrections to the partial
decay widths of the Higgs bosons decaying into a pair of DM particles.  
Hereafter, we regard the CP-even dark scalar $H_{D}$ as the DM
candidate unless  otherwise specified. We will therefore present the  
analytic expressions for the decay widths of $H_{i}\to H_{D}H_{D}$
$(i=1,2)$ at NLO. \s

The decay rate for $H_{i}\to H_{D}H_{D}$ at LO is given  by
\begin{align}
\Gamma^{\rm LO}(H_{i}\to
  H_{D}H_{D})=\frac{1}{32\pi^{2}m_{H_{i}}}\lambda_{H_{i}H_{D}H_{D}}^{2}\sqrt{1-\frac{4m_{H_{D}}^{2}}{m_{H_{i}}^{2} 
  } }\,,   
\end{align}
where the scalar coupling $\lambda_{H_{i} H_{D} H_{D}}$ is given in Eq.~\eqref{eq:lam1DD}.
 The 1PI diagrams contributing to the
 one-loop amplitude  for the process $H_{i}\to H_{D}H_{D}$ contain UV
 divergences 
 that are absorbed by introducing the corresponding counterterms in the amplitude. 
Shifting all parameters in Eq.~\eqref{eq:lam1DD}, we obtain the counterterms for the $\lambda_{H_{i}H_{D}H_{D}}$ couplings,
\begin{align}
\notag
\delta \lambda_{H_{i}H_{D}H_{D}}^{\rm para.}&=-2\Big[
\frac{R_{i1}}{v}(\delta m_{H_{D}}^{2}-\delta m_{22}^{2} )
+\frac{1}{v}(m_{H_{D}}^{2}-m_{22}^{2}-\frac{1}{2}v_{S}^{2}\lambda_{8}	)\delta R_{i1}\\ \notag
&+\frac{v_{S}}{2}\lambda_{8}\delta R_{i3}+\frac{1}{2}\frac{v_{S}}{v}(R_{i3}v-R_{i1} v_{S})\delta \lambda_{8}
+\frac{R_{i1}}{v^{2}}(m_{22}^{2}-m_{H_{D}}^{2}+\frac{1}{2}v_{S}^{2}\lambda_{8} )\Delta v \\ 
&+(\frac{R_{i3}}{2}-R_{i1}\frac{v_{S}}{v})\lambda_{8}\Delta v_{S}
\Big].
\end{align}
The counterterms $\delta R_{i1}$ and $\delta R_{i3}$ are those of the 3 $\times$ 3 mixing matrix for the neutral Higgs bosons, $R$ (Eq.~\eqref{eq:NH}). 
For instance,  when $i=1$, we obtain
\begin{align}
\delta R_{11}&=\delta c_{\alpha}=-s_{\alpha}\delta \alpha\,, \\ 
\ \ \ \delta R_{13}&=\delta s_{\alpha}=c_{\alpha}\delta \alpha\,. 
\end{align}
As previously discussed we have three options for the counterterms $\delta \lambda_{8}$ and
$\delta m_{22}^{2}$, namely, the $\overline{\rm
  MS}$ scheme, the OS process-dependent scheme
and the ZEM process-dependent scheme. 
The corresponding conditions and counterterms are given in
Eq.~\eqref{eq:CTlam8m22MS}, Eq.~\eqref{eq:PDS} and
Eq.\eqref{eq:newPDS}, respectively, together with
Appendix~\ref{ap:lam8m22} .  
In addition, performing the shift of the fields present in the tree-level Lagrangian for the $H_{i}H_{D}H_{D}$ vertices, we obtain
\begin{align}
\delta \lambda_{H_{i}H_{D}H_{D}}^{\rm field}&=\lambda_{H_{i}H_{D}H_{D}}\left(\delta Z_{H_{D}}+\frac{1}{2}\delta Z_{H_{i}}+\frac{1}{2}\frac{\lambda_{H_{j}H_{D}H_{D}}}{\lambda_{H_{i}H_{D}H_{D}}}\delta Z_{H_{j}H_{i}}\right),\ \ \ (j\neq i).
\end{align}
Therefore, the counterterms for the one-loop amplitudes for $H_{i}\to
H_{D}H_{D}$ are given by 
\begin{align}\label{eq:CTHHDHD}
\mathcal{M}_{H_{i}\to H_{D} H_{D}}^{\rm CT}=\delta \lambda_{H_{i}H_{D}H_{D}}^{\rm field}+\delta \lambda_{H_{i}H_{D}H_{D}}^{\rm para.}\,.
\end{align}
With this counterterm, the renormalized one-loop amplitude for $H_{i}\to H_{D} H_{D}$  is expressed as
\begin{align}
\label{eq:rHiHDHD}
\mathcal{M}_{H_{i}\to H_{D} H_{D}}^{\rm 1\mathchar`-loop}=\mathcal{M}_{H_{i}\to H_{D} H_{D}}^{\rm 1PI}+\mathcal{M}_{H_{i}\to H_{D} H_{D}}^{\rm CT}\,. 
\end{align}

We can finally write the decay width at NLO as
\begin{align}
\Gamma ^{\rm NLO} (H_{i}\to H_{D} H_{D})=\Gamma^{\rm LO}(1+\Delta^{\rm 1\mathchar`-loop})\,,
\end{align}
 where the one-loop corrections are written as
 \begin{align}
 \Delta^{\rm 1\mathchar`-loop}=2\frac{{\rm Re}(\mathcal{M}_{H_{i}\to H_{D} H_{D}}^{\rm 1\mathchar`-loop})}{\lambda_{H_{i}H_{D}H_{D}}}\,.
 \end{align}

\section{ Numerical Results}
\label{sec:num}
In this section, we analyze the impact of the one-loop corrections to the invisible decay of the SM-like Higgs boson. In Sec.~\ref{sec:NumericalWidth},
we start by discussing the
behavior of the corrections to the partial decay width of $H_{1}\to
H_{D}H_{D}$ with the most relevant parameters of the model, namely the trilinear tree-level coupling of the 125 GeV Higgs with the two DM
candidates and the mass difference between the two neutral scalars
from the dark sector.  
We then perform a scan in the allowed parameter space, in
Sec.~\ref{sec:NumericalBRs}, and present the results for the branching
ratios for the invisible decays of the Higgs bosons $H_i$ $(i=1,2)$.   
The calculations of the NLO corrections were performed using {\tt FeynRules~2.3.35}~\cite{Christensen:2008py,Degrande:2011ua,Alloul:2013bka},
 {\tt FeynArts~3.10}~\cite{Kublbeck:1990xc,Hahn:2000kx} and {\tt FeynCalc~9.3.1}~\cite{Mertig:1990an,Shtabovenko:2016sxi}. 
The same calculations were independently done using  {\tt SARAH~4.14.2}~\cite{Staub:2009bi,Staub:2010jh,Staub:2012pb,Staub:2013tta,Staub:2015kfa},  
{\tt FeynArts~3.10} and {\tt FormCalc~9.8}~\cite{Hahn:1998yk}. Loop
integrals were computed using {\tt LoopTools}~\cite{Hahn:1998yk,
  vanOldenborgh:1989wn}. 
We have checked numerically that the results obtained with the two
different procedures were in agreement. 
\subsection{Impact of the One-Loop Corrections on the Decay Rates}\label{sec:NumericalWidth}

We start by analysing our model  in the Inert Doublet Model (IDM)~\cite{Deshpande:1977rw}, which can be obtained as a limit of the DDP of the N2HDM
by setting (in this order) $\lambda_{8}=0,\ \alpha=0, \ v_{S}\to \infty $. The parameters chosen take into account the bounds for
the IDM presented in~\cite{Arcadi:2019lka}.
We will present numerical results for the one-loop corrected partial decay widths of the CP-even Higgs bosons to dark matter particles.
For a particular choice of parameters, we will compare the three
renormalization schemes for $\delta\lambda_8$ and $\delta  m_{22}^2$ showing the one-loop corrections in the 
$\overline{\rm MS}$ scheme, in the on-shell process-dependent scheme 
and in the ZEM process-dependent scheme. For
this comparison we are not taking into account any theoretical
constraints yet. The goal is to  
understand the theoretical behavior of the one-loop
corrections. Numerical results considering the theoretical constraints
as well as experimental constraints will be presented in the next section.  \s

Among the 13 free parameters given in Eq.~\eqref{eq:inpparaDDP}, 
the EW VEV is fixed by the
input parameters from the gauge sector, 
which we choose to be $m_{Z}$, $m_{W}$ and $\alpha_{G_\mu}$.  Using $\alpha_{G_\mu}$ allows us to resum large logarithms from the
light fermion contributions. In this sense, our result for
the decay width at LO does not correspond to the pure tree-level
result as a large universal
  part of the ${\cal O}(\alpha)$ corrections is already included at LO. 
%
   \begin{figure}[b!] \centering
    \hspace{0.5cm}
        \includegraphics[width=70mm]{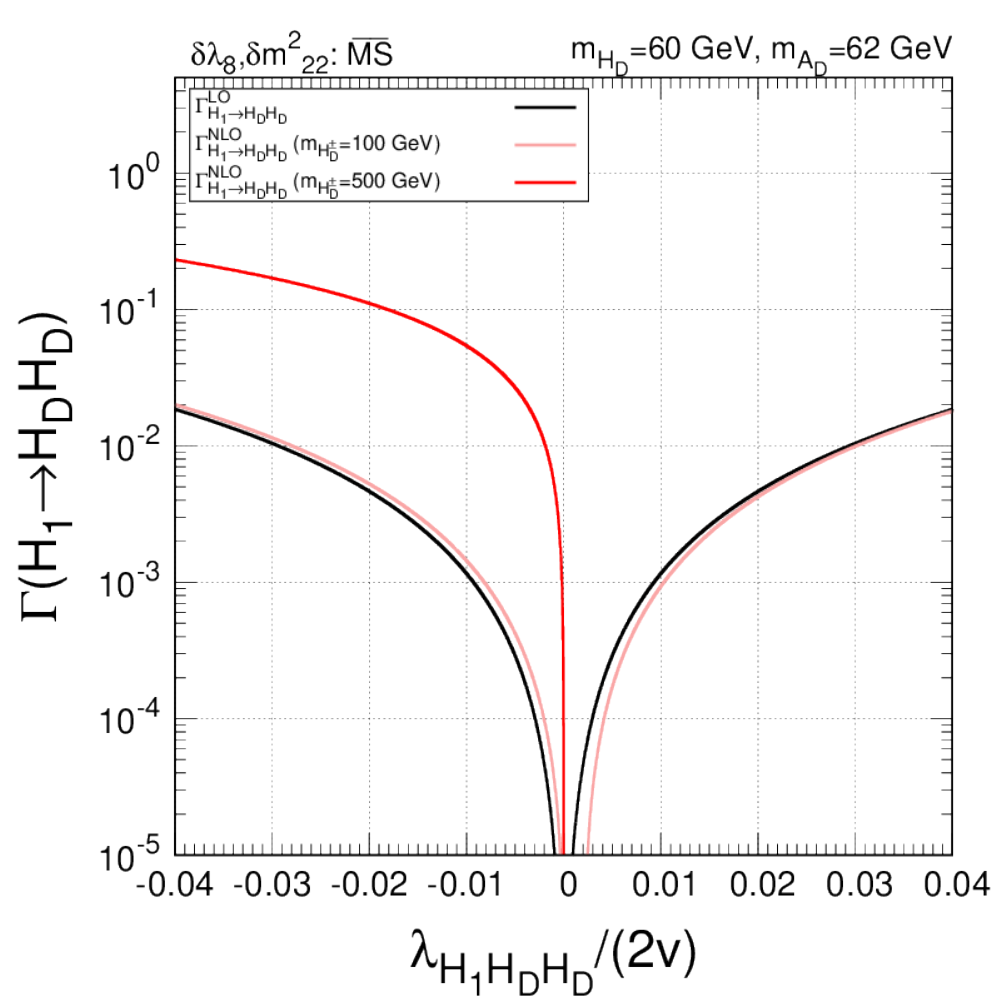} \hspace{1cm}
        \includegraphics[width=70mm]{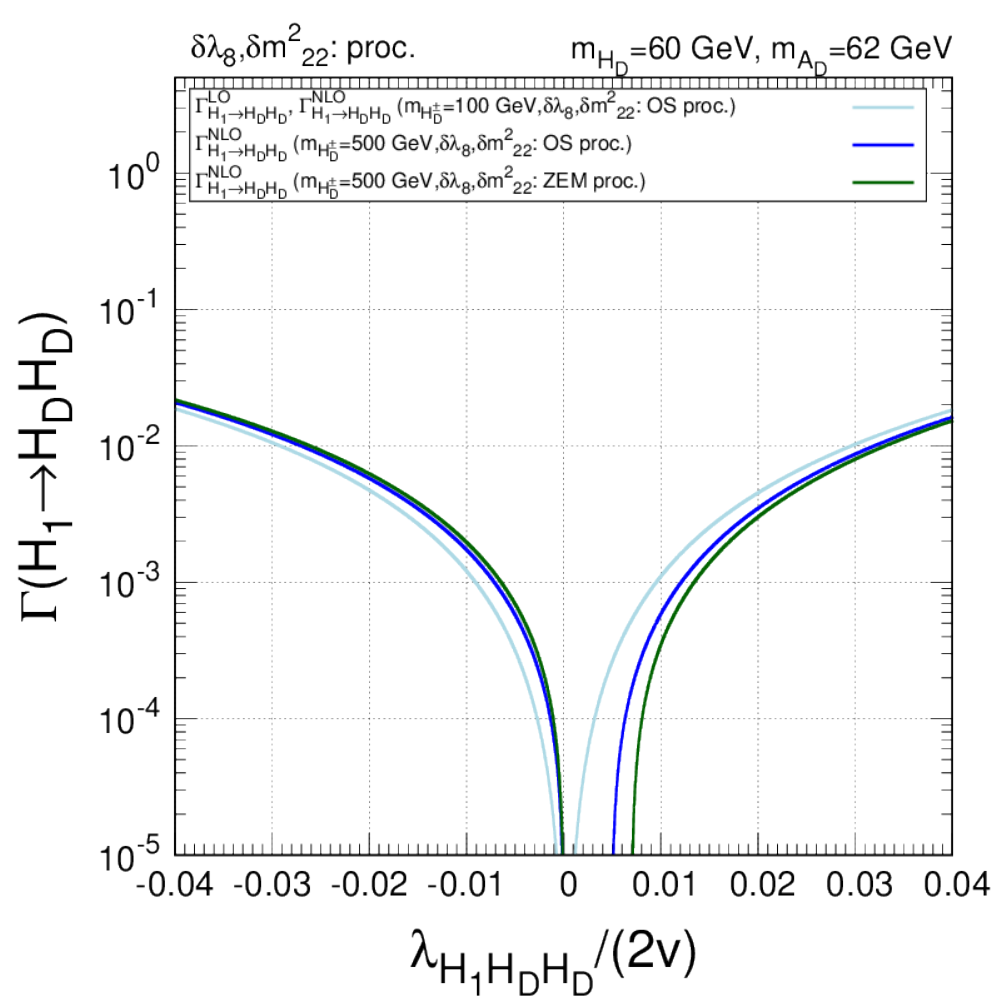}    
     \vspace{0cm}
      \caption{Decay width $H_{1}\to H_{D}H_{D}$ at LO and NLO as a
        function of the tree-level coupling
        $\lambda_{H_{1}H_{D}H_{D}}$ for $m_{H_{D}}=60$~GeV and
        $m_{A_{D}}=62$~GeV, and for two different values of 
        $m_{H_D^\pm}$, $m_{H_D^\pm}=100$ and 500 GeV. The variable $m_{22}^{2}$
      is scanned in a range such that the DM direct detection constraints
      hold. Other input parameters are fixed to the values given in
      Eqs.~\eqref{eq:m2fix}, \eqref{eq:BPFig1and2} and \eqref{eq:IDMlimit}. The left
      panel shows the results for the  
      $\overline{\rm MS}$ scheme, while the right panel presents
      results for the two process-dependent schemes.  
      \label{fig:H1HDHDlambda} 
}\end{figure}
The remaining 10 parameters, besides the two tadpoles
  $T_\Phi=T_S=0$, are set as follows: $H_{1}$ is the SM-like Higgs
boson with $m_{H_{1}}=125.09$ GeV, and the mass of the heavier Higgs
boson $H_2$ is fixed as 
\beq 
m_{H_{2}}=500 \, {\rm GeV} \;. \label{eq:m2fix}
\eeq  
The parameters of the dark sector, $m_{H_{D}}$, $m_{H^{\pm}_{D}}$ and
$\lambda_{2}$, are set to
\begin{align}\label{eq:BPFig1and2}
m_{H_{D}}=50\ \mbox{GeV or }60~{\rm GeV}\,,\quad
m_{H^{\pm}_{D}}=100\ \mbox{GeV or } 500\ \mbox{GeV} \,,\quad
\lambda_{2}=0.12\,,
\end{align}
while the remaining mass parameters $m_{A_{D}}$ and $m^{2}_{22}$ can be either scanned over or fixed in the following plots. 
We assume $m_{A_{D}}>m_{H_{D}}$, meaning that the dark scalar $H_{D}$ is the DM candidate. 
As previously stated we choose for $\lambda_{8}$, $\alpha$ and $v_{S}$, 
\begin{align}\label{eq:IDMlimit}
\lambda_{8}=0,\ \ \ \alpha=0,\ \ \ v_{S}\to \infty \,,
\end{align}
in that order, which is equivalent to take $m_{S}^{2}, \lambda_{6}, \lambda_{7}$ and $\lambda_{8}$ equal to zero in the scalar potential in Eq.~\eqref{eq:scalpot}. This is in turn
equivalent to the IDM potential. Hence, Eq.~\eqref{eq:IDMlimit} gives the IDM limit in the DDP phase of the N2HDM and we should recover the IDM results. 
When the $\overline{\rm MS}$ scheme is used to calculate $\delta
\lambda_{8}$ and $\delta m_{22}^{2}$, the one-loop amplitude for
$H_{1}\to H_{D}H_{D}$ depends on the renormalization scale $\mu$ and
we set it as $\mu^{2}=m_{H_1}^{2}$. \s 

In Fig.~\ref{fig:H1HDHDlambda}, we show the correlation between the tree-level coupling $H_{1} H_{D}H_{D}$ and the decay width for the corresponding
process $H_{1}\to H_{D}H_{D}$ at LO and NLO and for two different
  charged Higgs masses, $m_{H_D^\pm}=100$~GeV and 500 GeV.  
In this plot, we set the mass of the CP-odd dark scalar to $m_{A_{D}}=62$ GeV and vary $m^{2}_{22}$ in a range that forces the tree-level coupling to be  $|\lambda_{H_{1}H_{D}H_{D}}/(2v)|<0.05$ ~\cite{Arcadi:2019lka}.  
The upper bound for $|\lambda_{H_{1}H_{D}H_{D}}/(2v)|$ corresponds to the current bounds for direct detection of DM from XENON1T~\cite{Aprile:2018dbl}. 
From the left panel, in which the  $\overline{\rm MS}$ scheme results are shown, one can see a parabolic behaviour for the decay width at both LO and NLO,
with the width vanishing at $\lambda_{H_{1}H_{D}H_{D}}/(2v)=0$. 
The most important feature is that the NLO corrections strongly
  depend on the value of $m_{H_D^\pm}$ and can be very large even for
relatively small $\lambda_{H_{1}H_{D}H_{D}}$ if the mass of the dark
charged scalars is large ($m_{H_{D}^{\pm}}=500$ GeV in the plot).  
%
In the right panel of Fig.~\ref{fig:H1HDHDlambda}, results for the two
process-dependent schemes are shown.  
The behaviour of the results at NLO for $m_{H^{\pm}_{D}}=100 $ GeV is
similar for all three renormalization schemes. In particular, we have
confirmed that the result for the ZEM process-dependent scheme almost coincides
with that for the OS process-dependent scheme. However, the NLO
corrections for the decay width at $m_{H^{\pm}_{D}}=500 $ GeV  
are quite moderate in both process-dependent schemes, in contrast with
the $\overline{\rm MS}$ scheme. \s   

   \begin{figure}[h!] \centering
        \includegraphics[width=85mm]{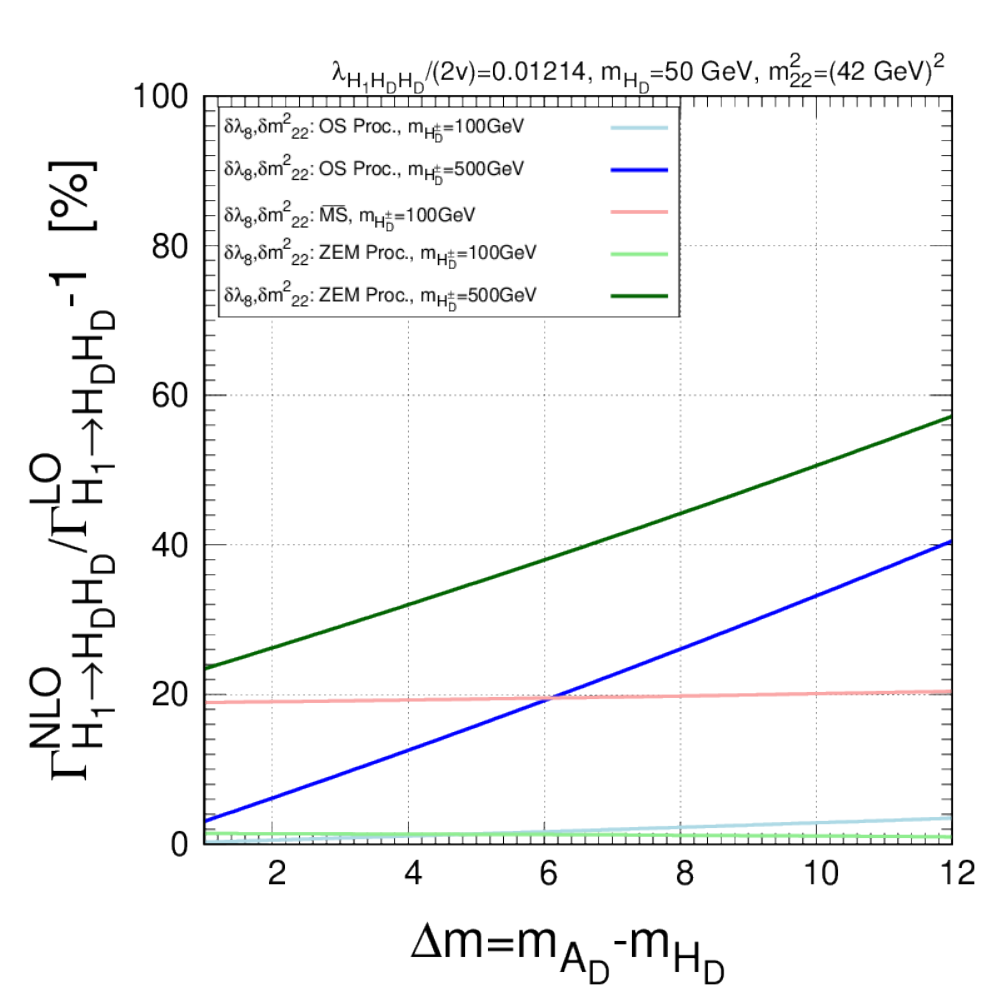}    
     \vspace{0cm}
      \caption{Relative size of the NLO corrections for $H_{1}\to
        H_{D}H_{D}$  as a function of the mass difference between
        $H_{D}$ and $A_{D}$, denoted by $\Delta
        m=m_{A_{D}}-m_{H_{D}}$, and for two different values of
        $m_{H^\pm_D}$, $m_{H_D^\pm}=100$ and 500 GeV.  
      The parameters are chosen as $m_{H_{D}}=50$ GeV and
      $m_{22}^{2}=(42\ {\rm GeV})^{2}$ while the mass of the CP-odd
      scalar $m_{A_{D}}$ is scanned keeping $m_{H_{D}}<m_{A_{D}}$.   
The other input parameters are fixed as in Eqs.~\eqref{eq:m2fix}, \eqref{eq:BPFig1and2}
and \eqref{eq:IDMlimit}. The line colours for the different schemes
are explained in the legend.   
      \label{fig:H1HDHDmdif2} 
      }
\end{figure}
In Fig.~\ref{fig:H1HDHDmdif2}, we show the relative size of the NLO
corrections
\beq
\Delta_{\text{NLO}}\equiv\Gamma^{\rm NLO}/\Gamma^{\rm LO}-1
\eeq
as a function of the mass difference between the CP-odd dark scalar
and the CP-even dark scalar, $\Delta m \equiv m_{A_{D}} - m_{H_{D}}$,
for the three different renormalization schemes and for two different 
  charged Higgs masses, $m_{H_D^\pm}=100$~GeV and 500 GeV.   
The parameters are chosen as $m_{H_{D}}=50~{\rm GeV}$ and $m_{22}^{2}= (42~{\rm GeV})^{2}$, which corresponds to $\lambda_{H_{1}H_{D}H_{D}}/(2v)=0.01214$. 
The upper limit $\Delta m \lesssim 12$ GeV  is used because we want to
compare the renormalization schemes in a region where they all can be applied. 
The SM-like Higgs decay into a pair of CP-odd scalars, $H_{1}\to
A_{D}A_{D}$, has to be kinematically allowed so that
Eq.~\eqref{eq:procreno} is applicable.  
The NLO corrections for the $\overline{\rm MS}$ scheme, with
$m_{H_{D}^{\pm}}$ fixed to 100 GeV, are almost constant, i.e., they do not depend on the mass difference
between the two dark neutral scalars. Nonetheless, as we have seen before,
they do depend quite strongly on the charged Higgs mass. 
In both process-dependent schemes, the NLO corrections strongly depend
on the mass difference, $\Delta m$, but also on the value of the charged
Higgs mass.  
For a low value of the charged Higgs mass, $m_{H^{\pm}_{D}}=100\,{\rm  
  GeV}$, the maximum value of the relative correction
for the process-dependent schemes is $\Delta_{\text{NLO}}$=4\%
 at $\Delta m=12~{\rm GeV}$, while the minimum is $\Delta_{\text{NLO}}\sim0\%$.  These corrections increase for larger charged Higgs mass.
 Considering  $m_{H^{\pm}_{D}}=500~{\rm GeV}$, the value of
 $\Gamma^{\rm NLO}/\Gamma^{\rm LO}-1$ has a minimum of about
 4\% (24\%) for the OS (ZEM) case for 
$\Delta m=0 $ and a maximum of about 40\% (57\%) for the OS (ZEM) 
case for $\Delta m=12~{\rm GeV}$. 
This behaviour can be understood from the fact that there is a significant number of terms in
$\mathcal{M}^{\rm  1\mathchar`- loop}_{H_{1}\to H_{D}H_{D} }$ that are proportional to
$\Delta m$ and, consequently, they have a large impact on the one-loop result.
The latter is also proportional to the charged Higgs mass and,
therefore, sizable corrections are found for $m_{H^{\pm}_{D}}=500$
GeV. In the $\overline{\rm MS}$ scheme, 
the NLO corrections for $m_{H^\pm}=500$~GeV are well
above 100\% in the entire mass range, $\Delta m\in [0,12]$~GeV, and
not shown in the plot. 

\subsection{Scan Analysis for the Branching Ratios }\label{sec:NumericalBRs}
In this section, we will perform a scan over the allowed
  parameter space of the model. This will enable us to understand the overall behavior of the NLO
corrections to the SM-like Higgs decays into a pair of DM particles. 
The evaluation of the branching ratio is performed using {\tt
  N2HDECAY}~\cite{Engeln:2018mbg} which is an extension of
the original code {\tt HDECAY}~\cite{Djouadi:1997yw,Djouadi:2018xqq} to the N2HDM.  
The program computes the branching ratios and the total decay widths
of the neutral Higgs bosons $H_{1}$ and $H_{2}$, including the
state-of-the art QCD corrections.  
Using the value of the partial widths evaluated by {\tt N2HDECAY},
$\Gamma^{\rm N2HDECAY}$, we evaluate the branching ratios for $H_{i}\to
H_{D}H_{D}$ with the NLO EW corrections as  

\begin{align}\label{eq:NLOBR}
{\rm BR}(H_{i}\to H_{D}H_{D})=\frac{\Gamma^{\rm \tt N2HDECAY}_{H_{i}\to
  H_{D}H_{D}}(1+\delta^{\rm EW}_{H_{i}\to H_{D}H_{D}}) }{\Gamma^{\rm
  \tt N2HDECAY}_{H_{i}\to \rm SM}+\Gamma_{H_{i}\to \Phi \Phi}  } \;,
\end{align}
where the correction factor $\delta^{\rm EW}_{H_{i}\to XX}$ is defined by
\begin{align}
\delta ^{\rm EW}_{H_{i}\to XX}=\frac{\Gamma^{\rm NLO}_{H_{i}\to XX}-\Gamma^{\rm LO}_{H_{i}\to XX}}{\Gamma^{\rm LO}_{H_{i}\to XX}}\, .
\end{align}
In Eq.~\eqref{eq:NLOBR}, the total decay width is separated into the
decays into the SM particles, $\Gamma^{\rm \tt N2HDECAY}_{H_{i} \to
  {\rm SM} }$, and the decay into a pair of the scalar bosons
$\Gamma_{H_{i} \to \Phi \Phi}$,  
defined as
\begin{align}
\Gamma^{}_{H_{1} \to  \Phi \Phi} &= \Gamma^{\rm \tt N2HDECAY}_{H_{1} \to  H_{D}H_{D} }\left(1+\delta^{\rm EW}_{H_{1}\to H_{D}H_{D}}\right)+\Gamma^{\rm \tt N2HDECAY}_{H_{1} \to  A_{D}A_{D} }\left(1+\delta^{\rm EW}_{H_{1}\to A_{D}A_{D}}\right)+\Gamma^{\rm \tt N2HDECAY}_{H_{1} \to  H^{+}_DH^{-}_D }\,, \\ 
\Gamma^{}_{H_{2} \to  \Phi \Phi} &= \Gamma^{\rm \tt N2HDECAY}_{H_{2} \to  H_{D}H_{D} }(1+\delta^{\rm EW}_{H_{2}\to H_{D}H_{D}})+\Gamma^{\rm \tt N2HDECAY}_{H_{2} \to  A_{D}A_{D} }\left(1+\delta^{\rm EW}_{H_{2}\to A_{D}A_{D}}\right)  \notag \\
&+\Gamma^{\rm \tt N2HDECAY}_{H_{2} \to  H_D^{+}H_D^{-} }+\Gamma^{\rm \tt N2HDECAY}_{H_{2} \to  H_{1}H_{1} }\,,
\end{align}
where we include our computed EW corrections to the decays into
neutral dark bosons,  
$H_{i}\to H_{D}H_{D}$ and $H_{i}\to A_{D}A_{D}$. Here we highlight that, in the process-dependent scheme,  $\delta^{\rm EW}_{H_{i}\to A_{D}A_{D}}$ disappears because of the renormalization condition Eq.~\eqref{eq:procreno}. \s


%

We consider two different scenarios in our scan.  
In {\it scenario 1}, the lighter Higgs boson $H_{1}$ is identified as
the SM-like Higgs boson and the other CP-even Higgs boson $H_{2}$ is
heavier than the SM-like Higgs boson. 
In {\it scenario 2},  $H_{2}$ is the SM-like Higgs boson and the other Higgs boson $H_{1}$ is lighter than the SM-like Higgs boson. 
In both scenarios, the dark scalar $H_{D}$ is the DM candidate. The scan is performed for the two scenarios to examine the impact of the NLO corrections 
in the allowed parameter space. We use for both scenarios the
following ranges for the parameters, 
\begin{align}
&1~{\rm GeV}<m_{H_{D}}<62 ~{\rm GeV}\,,\ \ \  
1~{\rm GeV}<m_{A_{D}}<1500 ~{\rm GeV}\, \, \, \,  (m_{A_D} > m_{H_D}) \notag \\
&65~{\rm GeV}<m_{H^{\pm}_{D}}<1500 ~{\rm GeV}\,, \ \ \
10^{-3}~{\rm GeV}^2<m_{22}^{2} < 5\cdot 10^{5}~{\rm GeV}^2\,,  \notag \\
& 1~{\rm GeV}<v_{S}<5000~{\rm GeV}\,,  \ \ \ 
- \pi/2 <\alpha< \pi/2\,,   \notag \\
&0<\lambda_{2}< 4 \pi\, ,\ \ \ 
- 4 \pi <\lambda_{8}< 4 \pi\,.
\end {align}
We chose $m_{H_D^\pm}$ to be above 65 GeV to prevent the SM-like
Higgs boson decay into a pair of charged Higgs
particles. Additionally, $\lambda_2$ is set positive due to the
boundedness from below (BFB) conditions,
see~\cite{Muhlleitner:2016mzt} for details on BFB of the N2HDM. \s

In scenario 1, the masses of the CP-even Higgs bosons are set  as
\begin{align}
m_{H_{1}}=125.09~{\rm GeV},\quad
130~{\rm GeV}<m_{H_{2}}<1500~{\rm GeV}. 
\end {align}
In scenario 2, they are taken as
\begin{align}
1~{\rm GeV}< m_{H_{1}}<120~{\rm GeV}, \quad  m_{H_{2}}=125.09~{\rm GeV}.
\end {align}
Since we focus on the case where ${H_{D}}$ is the DM particle, we
assume $m_{H_{D}}<m_{A_{D}}$.  \s

\begin{figure}[h!]\centering
          \includegraphics[width=75mm]{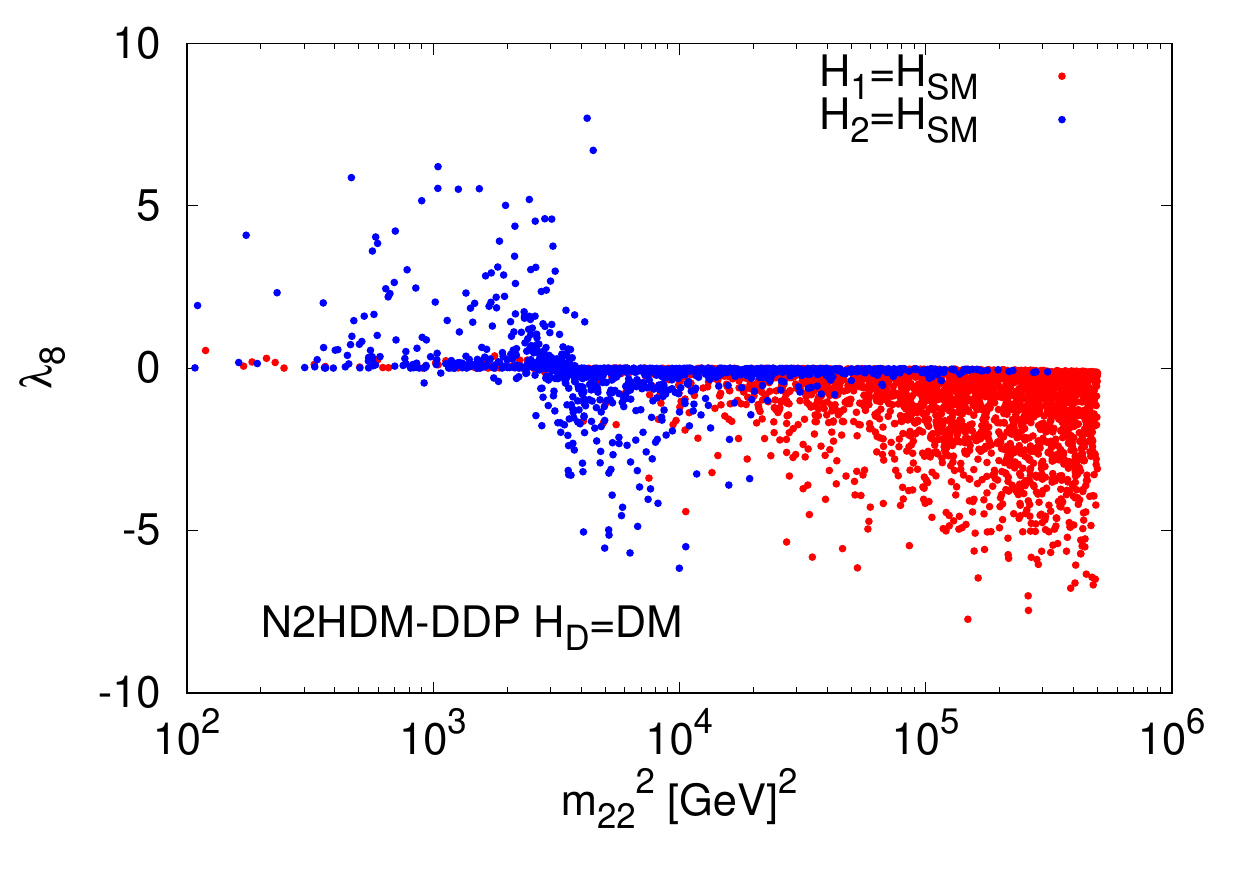}
          \includegraphics[width=75mm]{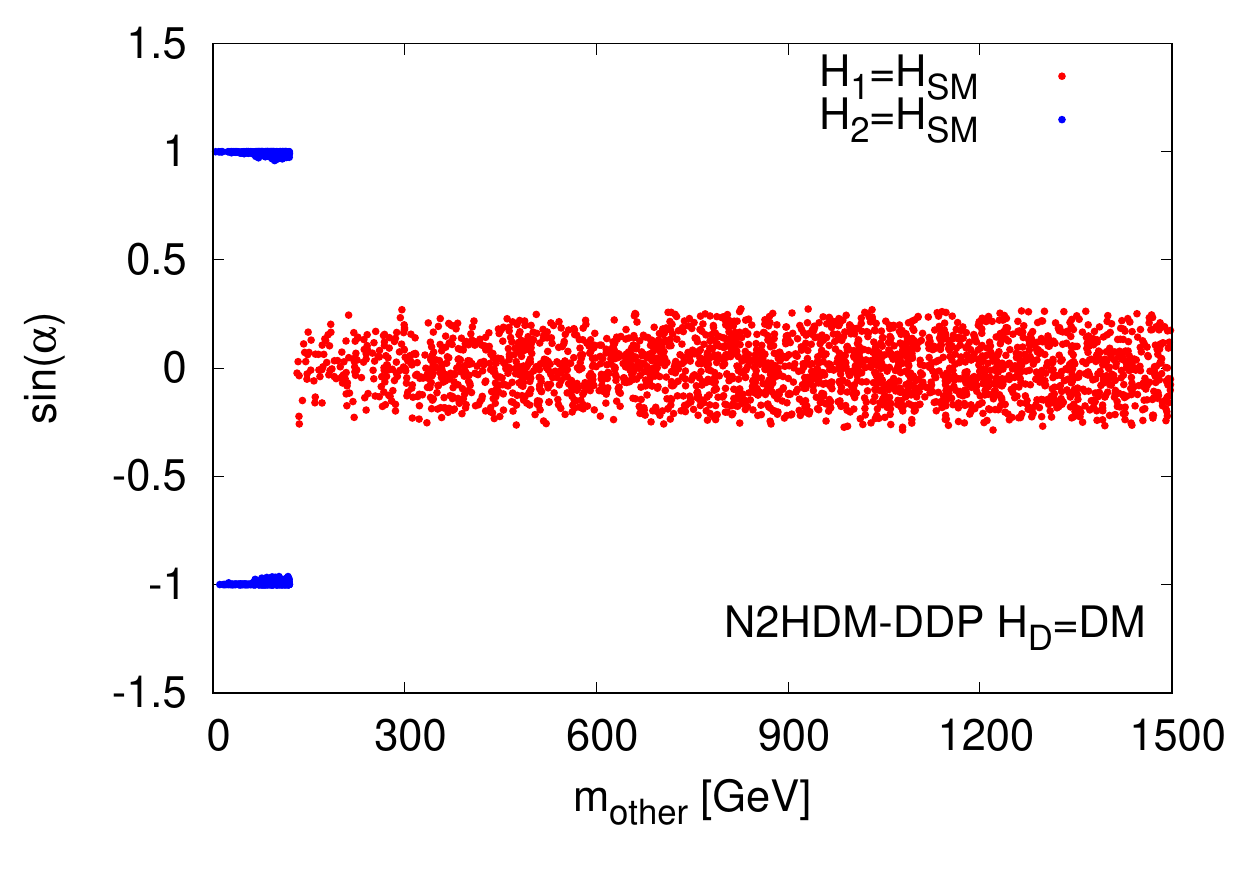}\\
   \caption{Projections of the allowed parameter space in the planes
     $(\lambda_8, \, m_{22}^2)$ (left) and $(\sin \alpha, m_{\text{other}})$
     (right), where $m_{\text{other}}$ is the mass of the non-SM-like Higgs
     boson. 
     The red points are for scenario 1 and the blue points are for
     scenario 2. 
 \label{fig:parspa}
     }
\end{figure}

Using {\tt ScannerS}~\cite{Coimbra:2013qq, Muhlleitner:2020wwk}, we generate input parameter points that pass the most relevant theoretical and experimental constraints. 
For the theoretical constraints~\cite{Muhlleitner:2016mzt,
  Engeln:2020fld}, {\tt ScannerS} evaluates perturbative unitarity,
boundedness from below and vacuum stability. 
The following experimental constraints are taken into account:
electroweak precision data, Higgs
 measurements, Higgs exclusion limits, and DM constraints.  
These constraints are included in {\tt ScannerS} via the interface
with other high energy physics codes:  {\tt
  HiggsBounds-5}~\cite{Bechtle:2020pkv} for the Higgs searches and
{\tt HiggsSignals-2}~\cite{Bechtle:2020uwn}  
for the constraints of the SM-like Higgs boson measurements.
For the DM constraints, the relic abundance and the nucleon-DM  cross
section for direct detection are calculated by {\tt
  MicroOMEGAs-5.2.4}~\cite{Belanger:2006is,Belanger:2008sj,Belanger:2018mqt}. The
DM relic abundance has to be below the value measured by the Plank
experiment~\cite{Aghanim:2018eyx}  and  the DM-nucleon
cross section has to be within the bounds imposed by the
XENON1T~\cite{Aprile:2018dbl} results. All points presented in the
plots have passed all the above constraints. In Fig.~\ref{fig:parspa} 
we show two projections of the allowed parameter space in the planes $(\lambda_8, \, m_{22}^2)$ (left) and $(\sin \alpha, m_{\text{other}})$ (right), where $m_{\text{other}}$ is the mass of the non-SM-like Higgs boson.
\begin{figure}[b!]\centering
          \includegraphics[width=70mm]{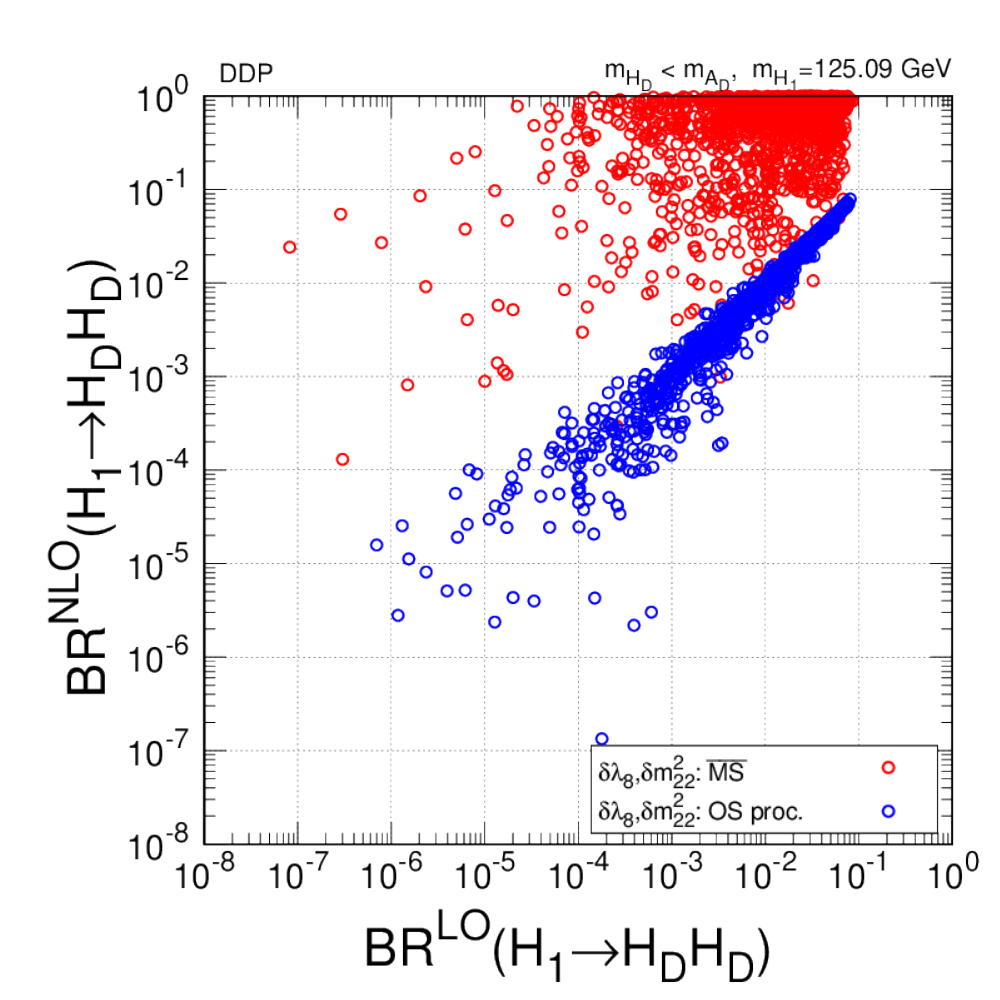}
          \includegraphics[width=70mm]{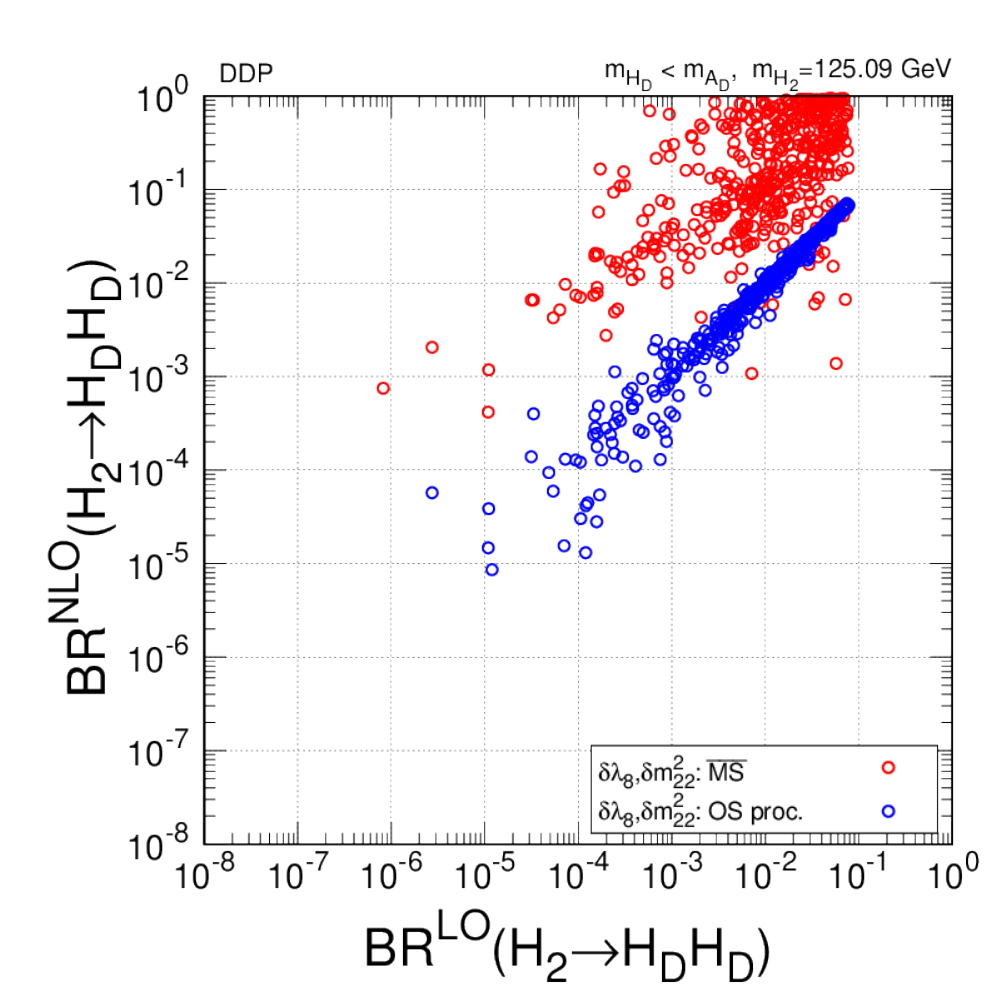}\\
   \caption{Correlation  between  the branching ratios at NLO and at LO in scenario 1 (left) and scenario 2 (right), respectively.   
   The red and blue points correspond to results in the $\overline{\rm
     MS}$ scheme and in the OS process-dependent scheme,
   respectively. 
  \label{fig:NLOLOBRs}
     }
\end{figure}
%
The red points are for scenario 1 and the blue points are for scenario
2. There are no particularly important features in the parameters
$\lambda_8$ and $m_{22}^2$ that probe the dark sector  
as expected, except for theoretical constraints that limit the quartic
couplings. As for  $\sin \alpha$, due to the  very SM-like behaviour of the discovered Higgs boson, $\sin \alpha$  is either 
close to zero or close to $\pm$1, depending on the considered scenario. \s

In Fig.~\ref{fig:NLOLOBRs}, we show the correlation between the 
BR$({H_{i} \to H_{D}H_{D}})$ calculated at LO and at NLO in scenario 1
(left panel) and in scenario 2 (right panel). 
The red and blue points correspond to the calculations in the
$\overline{\rm MS}$ scheme and in the OS process-dependent scheme,
respectively.  This sample has points with $m_{A_D} < 125/2$ GeV.
The first important thing to note is that in both scenarios the LO BR is always below 10\%. 
The main reason for this to happen is the very precise measurements of the Higgs couplings to SM particles which indirectly limit the Higgs
coupling to new particles. \s

The NLO corrections have a very different behaviour in the two renormalization schemes presented.
For the $\overline{\rm MS}$ scheme, the NLO corrections are not reliable with NLO BRs reaching 100\% in both scenarios.
Conversely, the OS process-dependent scheme is better behaved. 
This behaviour, in the OS scheme, can be traced back to the suppression of the NLO corrections by the mass difference between $H_{D}$ and $A_{D}$, 
as explained in Sec.~\ref{sec:NumericalWidth}.  
In our analysis, the mass difference  is in the range  $0~{\rm GeV}\lesssim\Delta m \lesssim 6~{\rm GeV}$
which leads to small corrections in the OS process-dependent scheme. \s


\begin{figure}[h!]\centering
          \includegraphics[width=75mm]{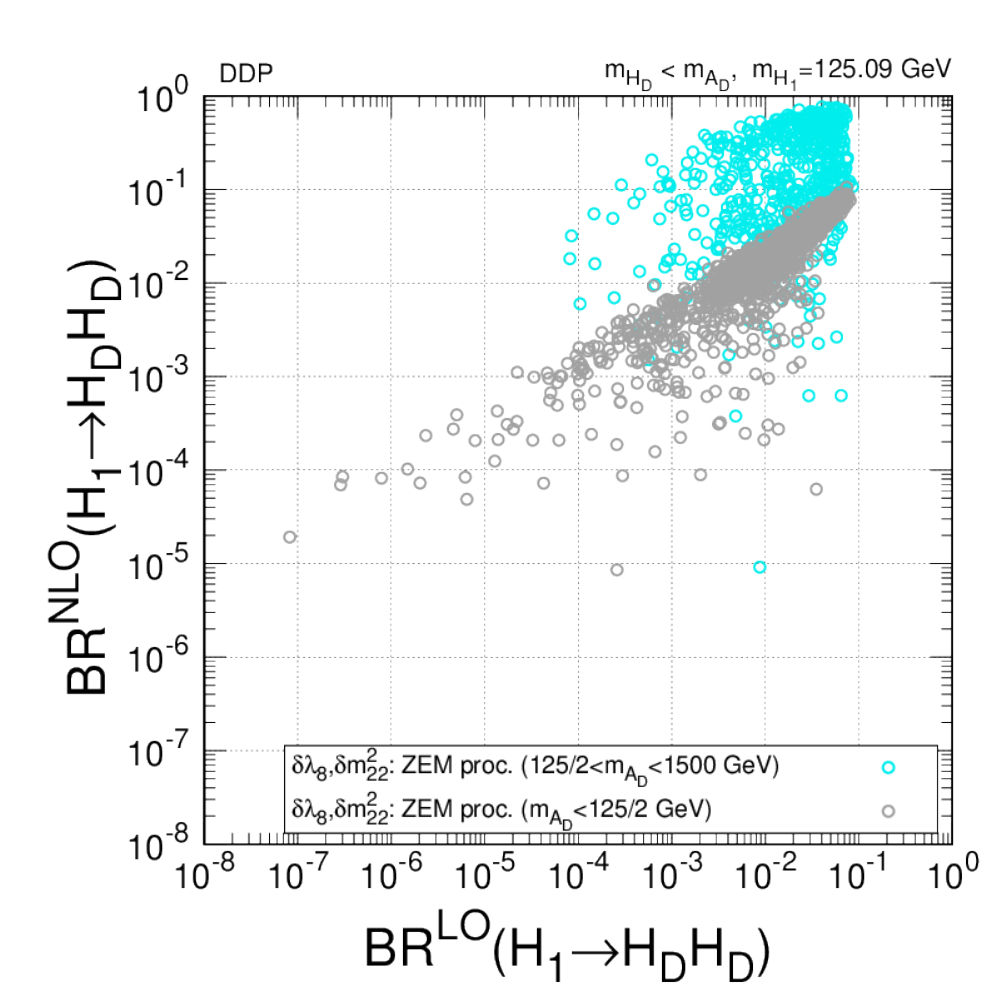}
          \includegraphics[width=75mm]{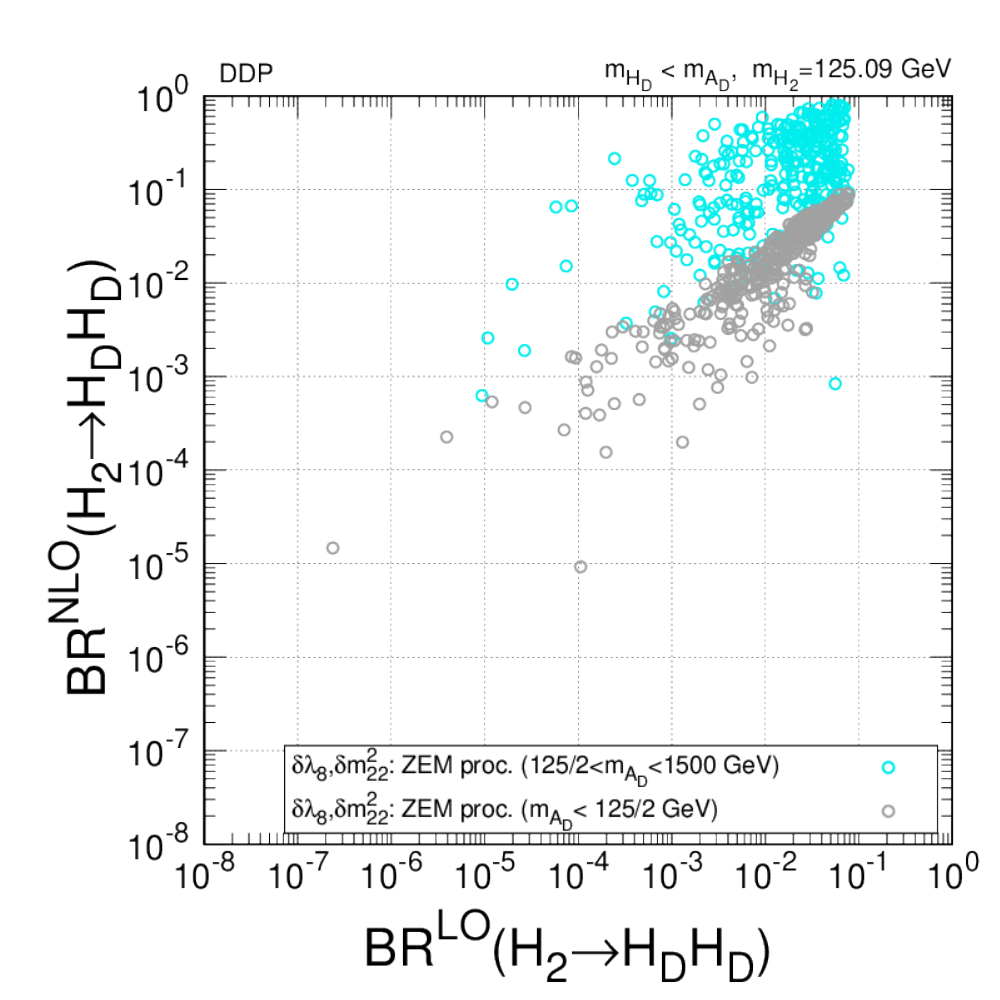}\\
   \caption{Correlation  between  the branching ratios at NLO and at LO in scenario 1 (left) and scenario 2 (right).  
   All points have been obtained in the ZEM process-dependent
   scheme. The grey points correspond to the previous sample where
   $m_{A_D} < 125/2$ GeV, while the blue points correspond to a range
   that is only allowed in the ZEM 
scheme, $125/2$ GeV$ <m_{A_D} < 1500$ GeV. }
  \label{fig:NLOLOBRsnew}
     
\end{figure}
In Fig.~\ref{fig:NLOLOBRsnew}, we show results for the ZEM process-dependent scheme. Again we display the correlation  between  the branching ratios at NLO and at LO in scenario 1 (left panel) and scenario 2 (right panel).  
We show results for two different samples of points, all calculated in the ZEM scheme. The grey points correspond to the previous sample where $m_{A_D} < 125/2$ GeV, while
the blue points correspond to a range that is only allowed in the ZEM scheme, $125/2$ GeV$ <m_{A_D} < 1500$ GeV. 
The points for which $m_{A_D} < 125/2$ have an overall similar behaviour as the ones for the OS scheme, in the sense that the NLO BRs are all below 0.1.
However, one can see that the corrections are much larger, even
for this sample. When we look at the blue points the picture
changes radically. This clearly shows that when the mass difference
between $H_D$ and $A_D$ is large the corrections become unstable. \s

\begin{figure}[h!]\centering
            \includegraphics[width=75mm]{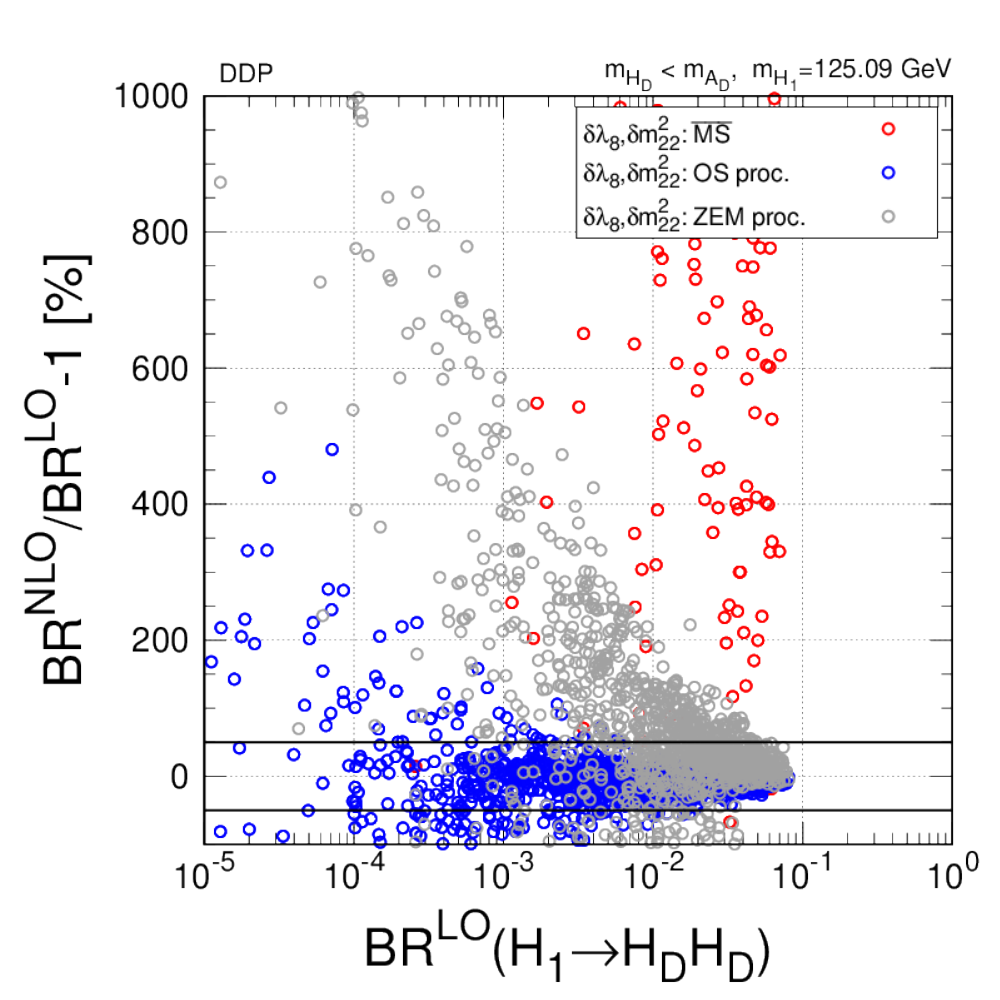}
          \includegraphics[width=75mm]{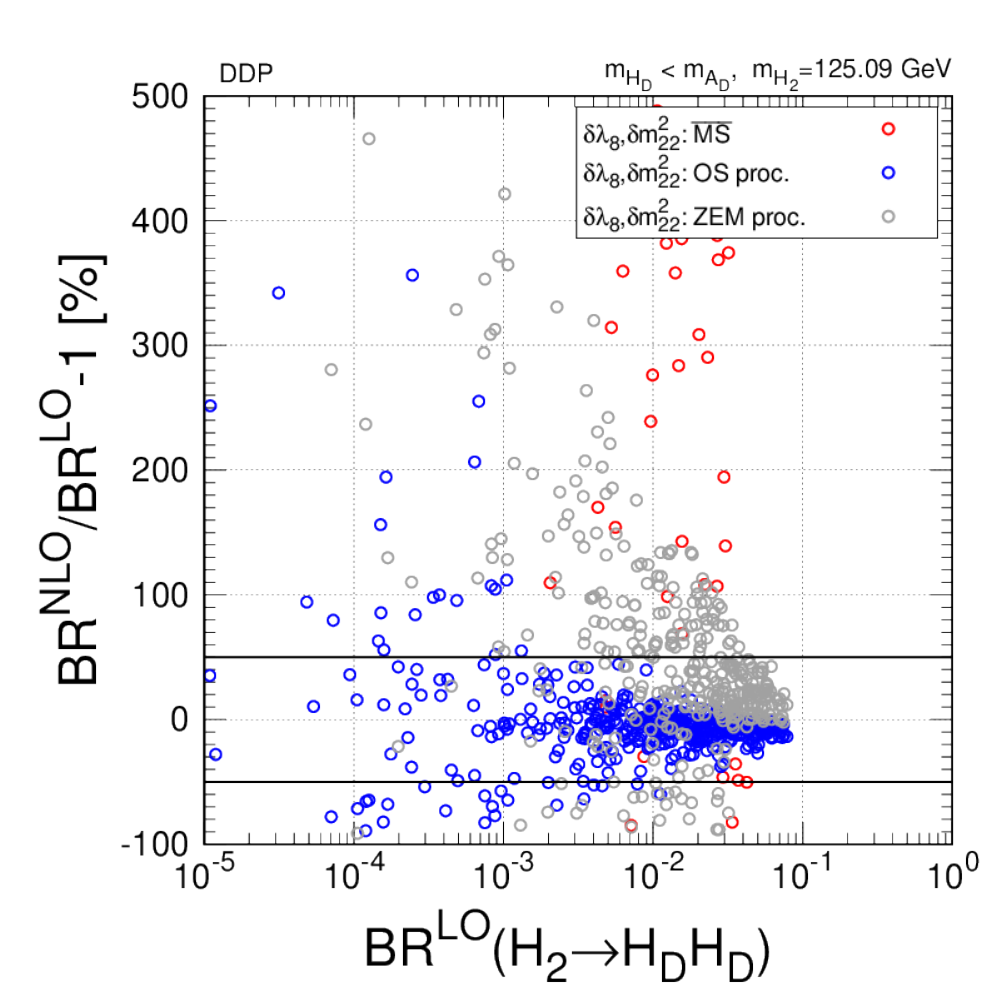}
   \caption{Ratio of NLO to LO corrections for ${\rm BR}(H_{i}\to H_{D}
       H_{D} )$ in scenario 1 (left) and scenario 2  (right). 
   The red, blue and grey points correspond to the $\overline{\rm
     MS}$, ZEM process-dependent and OS process-dependent scheme, respectively. 
   The black horizontal lines correspond to  ${\rm BR}^{\rm NLO}/{\rm BR}^{\rm LO}-1=\pm 50\%$. }
  \label{fig:LONLOkfac}
     
\end{figure}

In order to understand to what extend these corrections depend on the
  renormalization schemes, we show, in Fig.~\ref{fig:LONLOkfac}, the ratio of NLO to LO corrections for the 
processes ${\rm BR}(H_{1}\to H_{D} H_{D} )$ (left panel) and  ${\rm BR}(H_{2}\to H_{D} H_{D} )$  (right panel). 
Here again the sample used is the one where $m_{A_D} < 125/2$ GeV.
The red, blue and grey points correspond to the $\overline{\rm MS}$,
OS process-dependent and ZEM process-dependent renormalization schemes, respectively. 
The black horizontal lines corresponds to  ${\rm BR}^{\rm NLO}/{\rm
  BR}^{\rm LO}-1=\pm 50\%$. The plots clearly show that the OS
process-dependent scheme is more stable with most corrections between
-50\% and 50\%. In any case, the corrections in this scheme can still
go up to 480\%. As the corrections above 100\% only occur for small
values of the LO BRs, 
the NLO values of the BRs are still well below the experimental
bound. The other two schemes are less stable and this is particularly
true for the $\overline{\rm MS}$ scheme. \s 

\begin{figure}[h!]\centering
            \includegraphics[width=75mm]{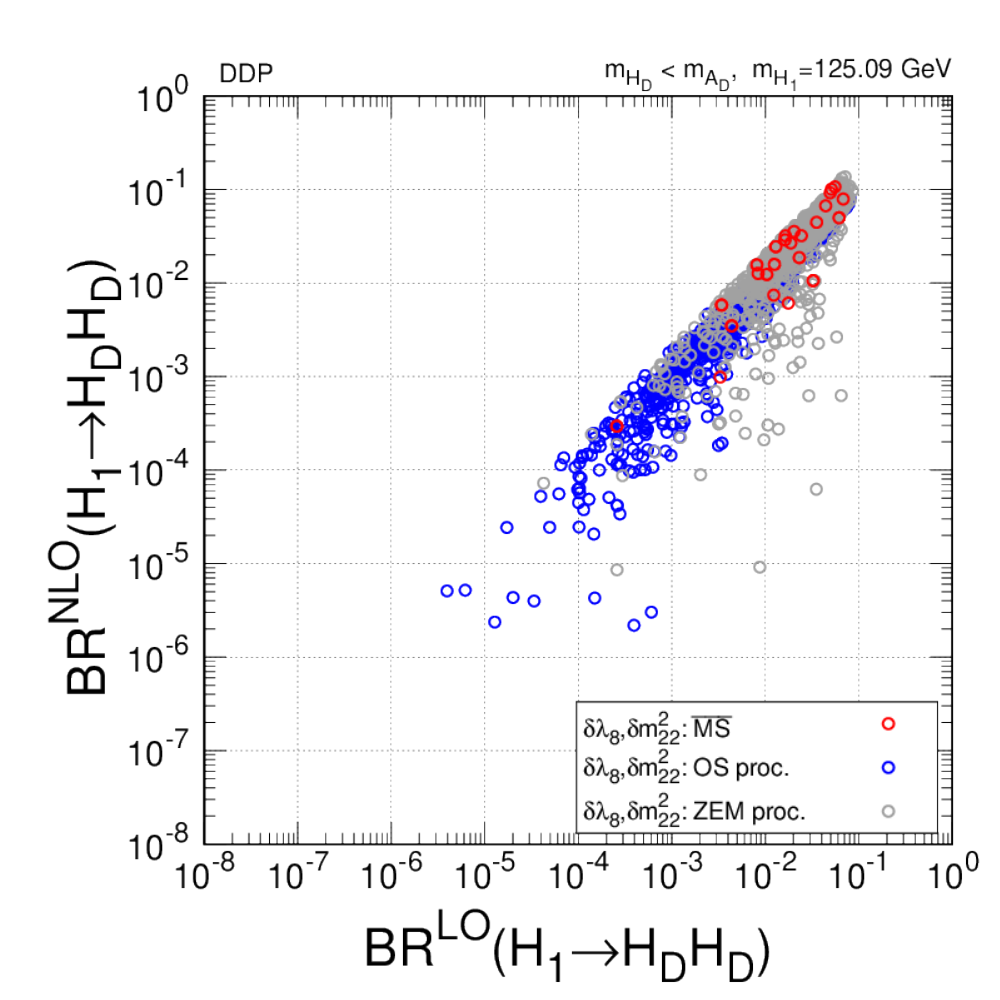}
          \includegraphics[width=75mm]{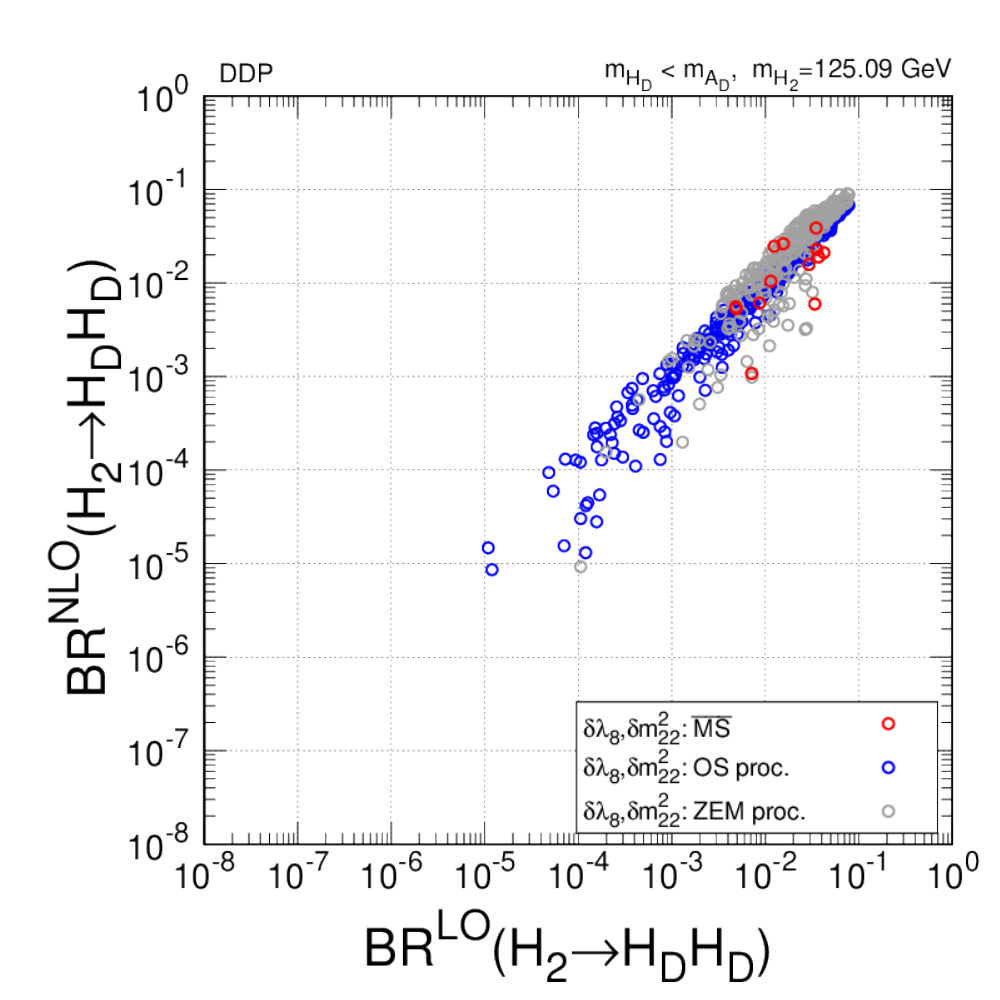}
   \caption{ Correlation  between  the branching ratios at NLO and at LO in scenario 1 (left) and scenario 2 (right), respectively. 
   The red, blue and grey points correspond to the $\overline{\rm
     MS}$, OS process-dependent and ZEM process-dependent scheme,
   respectively.}
  \label{fig:LONLOkfacnew}
     
\end{figure}

A clearer picture of the results for the NLO
corrections that can be trusted in terms of perturbation theory can be achieved by considering only the points for which the corrections are below 100\%.
In Fig.~\ref{fig:LONLOkfacnew}, we present the correlation  between  the branching ratios at NLO and at LO in scenario 1 (left panel) and scenario 2 (right panel), respectively. 
All points presented have NLO corrections below 100\%  and all points with NLO corrections above 100\% were discarded. We conclude that the surviving points are all still below
the current experimental limit for the Higgs invisible BR apart
  from a few grey points. One should keep in mind the theoretical
  uncertainties due to missing higher-order corrections. Additionally,
  the other decay channels do not include electroweak corrections,
  which has an impact on the branching ratios. Given these
  caveats, the points can still be considered compatible with the
  experimental results.\footnote{While the loop-corrected decay width
    would clearly exemplify the effect of the corrections, we still show
    the branching ratios to get an approximate estimate of the
    compatibility of the model with the experimental 
    results. We will compute and include the EW corrections to all
    decay processes in future work.} 
Therefore, no constraints on the parameter space come from including
the NLO corrections. However, as the limit  
on the Higgs invisible BR improves, there are now ranges of allowed values for the NLO corrections that will certainly lead to constraints on the parameter space.
\s

We end this section with a comment about the case where  $A_{D}$ is the DM candidate. We have also calculated the NLO corrections to Higgs boson invisible decays in the case that $A_{D}$ is the DM particle.    
The renormalization was done using the process dependent scheme with
the decay $H_{i}\to H_{D}H_{D}$ ($i=1,2$), i.e., $\Gamma_{H_{i}\to H_{D} H_{D}}^{\rm LO}\overset{!}{=} \Gamma_{H_{i}\to H_{D} H_{D}}^{\rm NLO} $ and 
we have performed the same scan analysis presented above for $H_D$. 
We confirm that the results are virtually identical with those
obtained for $H_D$, in both scenarios 1 and 2.

\section{Conclusions} 
\label{sec:conc}

In this work we have calculated the EW NLO corrections to the branching ratio of the SM-like Higgs boson invisible decay in the DDP of the N2HDM.
We have analysed two different scenarios, one where the SM-like Higgs
is the lighter of the visible two CP-even scalars and one where
it is the  heavier. There are, however, no significant differences between the two scenarios.
The model has 13 input parameters from the scalar
sector. Masses and wave function renormalization constants are renormalized on-shell. 
The rotation angle is renormalized by relating the fields in the gauge basis and in the mass basis and it is ultimately defined with
the off-diagonal terms of the wave function renormalization
constants. We apply a gauge-independent renormalization scheme based
on the alternative tadpole scheme together with the pinch technique. Besides the EW VEV, that is renormalized exactly like in the SM, we still
have four parameters left: $\lambda_2$ (which does not enter in any of
the processes under study),  $\lambda_8$, $m_{22}^2$ and $v_S$. \s

Regarding $\lambda_8$ and $m_{22}^2$ the analysis was performed for three different renormalization schemes. 
The three schemes used for these parameters were the $\overline{\rm
  MS}$, the OS process-dependent and the ZEM
process-dependent scheme. Only in the first one does $v_S$ need to be renormalized.
The most stable scheme is the OS process-dependent one, but
it can still lead to corrections above 
100 \% in some regions of the parameter space. It should be noted that
one of the reasons for the OS scheme stability is that the mass
difference between the two neutral dark scalars, $m_{A_D} - m_{H_D}$,
is bounded to be below about 10 GeV. Note that the OS process-dependent
scheme needs an allowed on-shell decay of the SM-like Higgs boson to
both pairs of dark scalars. \s 

With the LHC run 3 starting soon, the Higgs coupling  and the
invisible Higgs decay width measurements will become increasingly
precise. It is clearly the time to understand what the NLO corrections
can tell us about the models with more precise measurements. In fact,
these experimental results can be the best,  
if not the only, tools available to probe the dark sectors postulated
as extensions of the SM. One should stress that the parameters
$\lambda_8$ and $m_{22}^2$ are only directly accessible 
through processes that involve the DM particles. The experimental
  sensitivity on the invisible decay width is now starting to become
  comparable to the limits imposed on the parameter space of the model
  from the coupling measurements. \s

We have found that the NLO corrections can be extremely large in some regions of the parameter space. Also, 
as we move to smaller values of the  ${\rm BR}(H_{i}\to H_{D} H_{D} )$, the corrections become larger and larger. This means
that the more constrained the BR is the more unstable are the NLO corrections. 
As a perturbativity criteria, we rejected all points for which the NLO corrections relative to the LO results are above 100\%.
With this condition, the behaviour of
NLO versus LO results is very much along the line BR$^{\text{NLO}} =$ BR$^{\text{LO}}$. Still, if the experimental bound of $BR(h \to$ invisible) improves, for instance to
 $10^{-2}$, the NLO result would vary between $\sim 10^{-4}$ to  $\sim 2 \times 10^{-2}$ .

\begin{appendix}
\section*{Appendix}
 \section{Determination of $\delta \lambda_{8}$  and $\delta m_{22}^{2}$ in the Process-Dependent Scheme}\label{ap:lam8m22}
In this appendix, we discuss how the counterterms $\delta \lambda_{8}$ and $\delta m^{2}_{22}$ are determined using the process-dependent scheme.
As previously discussed, although our starting point is to force the amplitudes at LO and at NLO to be equal, we choose
two different approaches. In one approach all particles are on-shell, which is equivalent to say that the condition is set
on the actual physical process. In the other approach, the condition is set at the amplitude level taking all external
momenta to be zero. The advantage of this second approach is not to
curtail the allowed parameter space. \s

We will now describe in detail the renormalization procedure for the on-shell case and discuss the differences when the external momenta are set to zero at the end of this appendix.
The on-shell process-dependent renormalization condition is to impose the decay widths for $H_{i}\to A_{D}A_{D}$ (i=1,2) calculated at NLO to be equal to the LO result, as expressed 
in Eq.~\eqref{eq:procreno} and repeated here for convenience,
\begin{align}
\Gamma_{H_{i}\to A_{D} A_{D}}^{\rm LO}
\overset{!}{=}
 \Gamma_{H_{i}\to A_{D} A_{D}}^{\rm NLO} \quad (i=1,2)\,.
\label{eq:condagain}
\end{align} 
This results in two equations where $\delta \lambda_{8}$ and $\delta m_{22}^{2}$ are the only unknowns. The remaining renormalization constants 
are all fixed.
Solving this set of equations, we can get expressions for these two counterterms.
The renormalization conditions, given in Eq.~\eqref{eq:condagain}, can be written as
\begin{align}\label{eq:RRCproc}
\frac{1}{2}\frac{1}{2m_{H_{i}}}\int d\Phi_{2}|\mathcal{M}_{i}^{\rm tree}|^{2}&=
\frac{1}{2}\frac{1}{2m_{H_{i}}}\int d\Phi_{2}\Big[|\mathcal{M}_{i}^{\rm tree}|^{2}+2{\rm Re}\big(\mathcal{M}_{i}^{{\rm tree}*}\mathcal{M}_{i}^{{\rm 1\mathchar`-loop}}\big)\Big] \notag \\
0&={\rm Re}\big(\mathcal{M}_{i}^{{\rm tree}*}\mathcal{M}_{i}^{{\rm 1\mathchar`-loop}}\big)\,,
\end{align}
where $d\Phi_2$ denotes the two-particle differential phase space volume and
$\mathcal{M}^{\rm tree/1\mathchar`-loop}_{i}\equiv \mathcal{M}^{\rm
  tree/1\mathchar`-loop}_{H_{i}\to A_{D}A_{D}}$ ($i=1,2$).  
The tree-level amplitude is given by 
\begin{align}
\mathcal{M}^{\rm tree}_{i}=\lambda_{H_{i}A_{D}A_{D}}\,,
\end{align}
 where  the scalar trilinear coupling $\lambda_{H_{i} A_{D}A_{D}}$ is
 \begin{align}
\label{eq:lamHADAD}
\lambda_{H_{i}A_{D}A_{D}}=-\frac{1}{v}\big[
2(m_{A_{D}}^{2}-m_{22}^{2})R_{i1}+\lambda_{8} v_{S}(vR_{i3}-v_{S}R_{i1})
\big]\,.
\end{align}
Taking into account that $\mathcal{M}_{i}^{{\rm tree}}$  is a real constant, the renormalization condition simplifies to  
\begin{align}\label{eq:1Loop}
 0&={\rm Re}\big(\mathcal{M}_{i}^{{\rm 1\mathchar`-loop}}\big)\,,
 \end{align}
with the one-loop amplitude expressed as
 \begin{align}
 \mathcal{M}_{i}^{{\rm 1\mathchar`-loop}}=\mathcal{M}_{i}^{\rm 1PI}
 +\mathcal{M}_{i}^{\rm CT}|_{\delta \lambda_{8},\  \delta m^{2}_{22}=0}
 +\mathcal{M}_{i}^{\rm CT}|_{\delta \lambda_{8}, \  \delta m^{2}_{22}\neq 0}\,,
 \end{align}
 where $\mathcal{M}_{i}^{\rm 1PI}$ denotes 1PI diagrams for the
   loop-corrected decay widths $H_{i}\to A_{D}A_{D}$ and the counterterm contributions are separated into $\delta \lambda_{8}$ and $\delta {m}_{22}^{2}$ dependent terms $\mathcal{M}_{i}^{\rm CT}|_{\delta \lambda_{8},\  \delta m^{2}_{22}\neq 0}$ and the remainder $\mathcal{M}_{i}^{\rm CT}|_{\delta \lambda_{8},\  \delta m^{2}_{22}=0}$.  
 We note that the counterterms for $H_{i}\to A_{D}A_{D}$ can be
 obtained from those for $H_{i}\to H_{D}H_{D}$, see Eq.~\eqref{eq:CTHHDHD},
 with the replacements
 
 \begin{equation} 
 (m_{H_{D}}, \lambda_{H_{i}H_{D}H_{D}},\delta
 m_{H_{D}}^{2},\delta Z_{H_{D}})\to (m_{A_{D}},
 \lambda_{H_{i}A_{D}A_{D}},\delta m_{A_{D}}^{2},\delta
 Z_{A_{D}}).
 \end{equation} 
Hence the counterterm amplitudes can be written as
\begin{align}
\mathcal{M}_{i}^{\rm CT}|_{\delta \lambda_{8}, \  \delta m^{2}_{22}\neq 0}&=
2\frac{R_{i1}}{v}\delta m^{2}_{22}-\frac{v_{S}}{v}(R_{i3}v-R_{i1}v_{S})\delta \lambda_{8}\,, \\
\mathcal{M}_{i}^{\rm CT}|_{\delta \lambda_{8}, \  \delta m^{2}_{22}= 0}&=-2\Big[
\frac{R_{i1}}{v}\delta m_{A_{D}}^{2}
+\frac{1}{v}(m_{A_{D}}^{2}-m_{22}^{2}-\frac{1}{2}v_{S}^{2}\lambda_{8}	)\delta R_{i1}+\frac{v_{S}}{2}\delta R_{i3}\notag \\
&+\frac{R_{i1}}{v^{2}}(m_{22}^{2}-m_{A_{D}}^{2}+\frac{1}{2}v_{S}^{2}\lambda_{8} )\Delta v 
+(\frac{R_{i3}}{2}-R_{i1}\frac{v_{S}}{v})\lambda_{8}\Delta v_{S}
\Big] \notag \\
&+\lambda_{H_{i}A_{D}A_{D}}\left(\delta Z_{A_{D}}+\frac{1}{2}\delta Z_{H_{i}}+\frac{1}{2}\frac{\lambda_{H_{j}A_{D}A_{D}}}{\lambda_{H_{i}A_{D}A_{D}}}\delta Z_{H_{j}H_{i}}\right),\ \ \ (j\neq i).
\end{align}
 Finally we obtain the following set of equations,
 \begin{align}\label{eq:CTlam8m22proc}
2\frac{R_{i1}}{v}\delta m^{2}_{22}-\frac{v_S}{v}(R_{i3}v-R_{i1}v_{S})\delta \lambda_{8}=-\mathcal{M}_{i}^{\rm 1PI}
 -\mathcal{M}_{i}^{\rm CT}|_{\delta \lambda_{8},\  \delta m^{2}_{22}=0}\,,
 \end{align}
which give us the expressions for $\delta \lambda_{8}$ and $\delta m^{2}_{22}$. 
Note that the left-handed side of Eq.~\eqref{eq:CTlam8m22proc} corresponds to the linear combinations of  $\delta \lambda_{8}$ and  $\delta m^{2}_{22}$ that also appear in the counterterms for $H_{i}\to H_{D} H_{D}$. \s

The second process-dependent scheme, where all external momenta are set to zero, also starts from the same set of Eq.~(\ref{eq:CTlam8m22proc}). The only difference is in the calculation of $\mathcal{M}_{i}^{\rm 1PI}$ in
 which the external momenta are set to zero instead of on-shell. \s 

 %
\begin{figure}[tbhp]\centering
            \includegraphics[width=70mm]{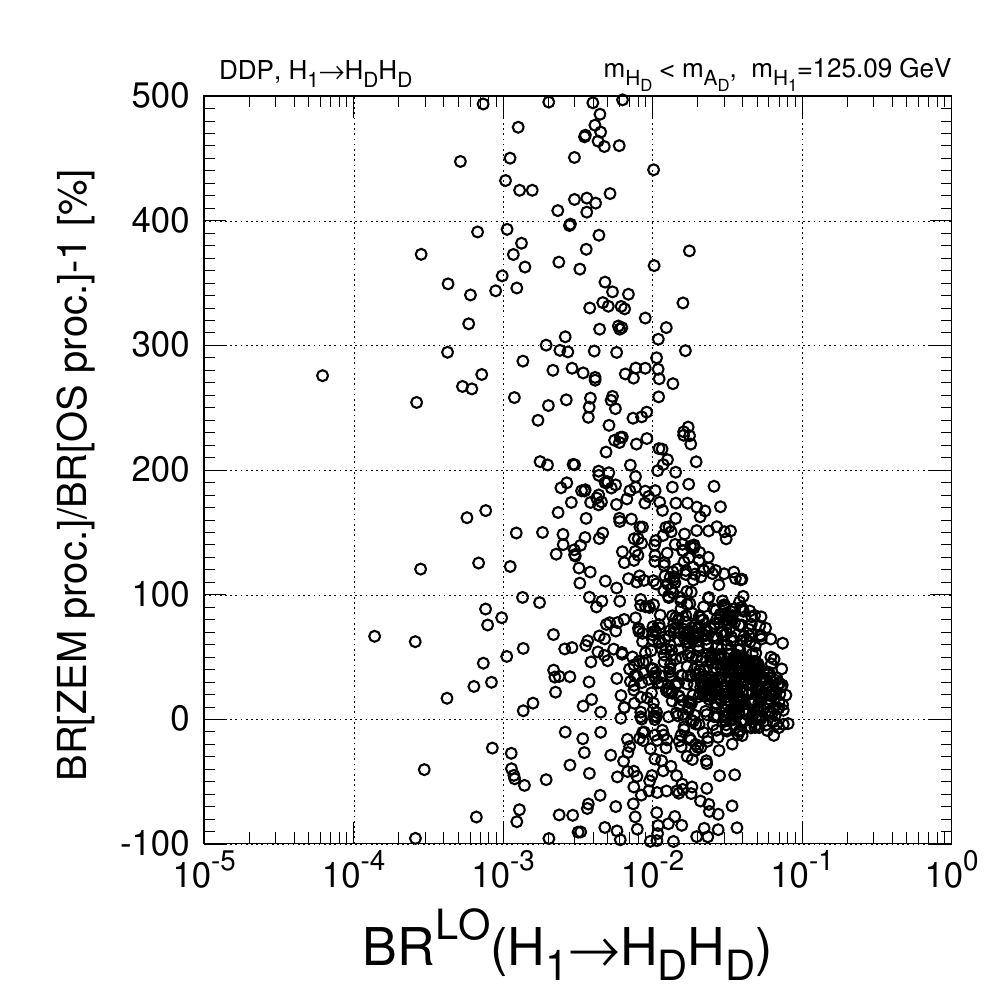}
          \includegraphics[width=70mm]{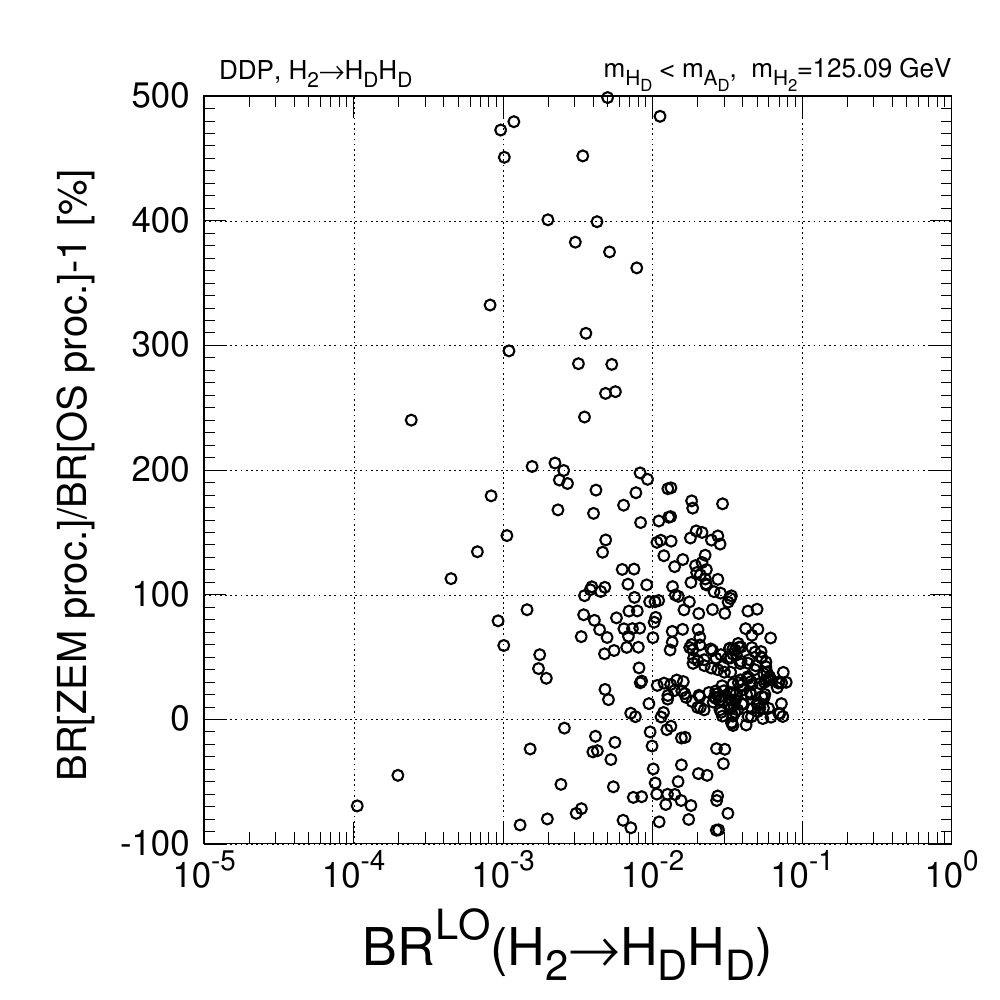}
   \caption{ Comparison of the two process-dependent renormalization schemes, on-shell vs. zero external momenta. We show the ratio between the NLO values vs. the corresponding 
   ${\rm BR}(H_{1}\to H_{D} H_{D} )$ (left) and ${\rm BR}(H_{2}\to
   H_{D} H_{D} )$ (right) at LO.}
  \label{fig:ratio}
\end{figure}
 %
 The two schemes are compared in Fig.~\ref{fig:ratio}, where we plot
 the ratio of the NLO corrections of the two process-dependent
 schemes, the zero external momenta over the on-shell scheme,
 in per-cent,
 as a function of the LO branching ratio. The left plot corresponds to
 the decay $H_{1}\to H_{D} H_{D}$ while the right one corresponds to
 $H_{2}\to H_{D} H_{D}$.  
 We conclude that the differences can be quite large. In fact, although we have cut the $y$-axis at 500 $\%$ for clarity, there  are points  where the corrections can go above $10^3 \%$,
 which, however, is not the case for the larger values of the LO branching
 ratios. The important point is that very large
 corrections only occur for the lower values of the BRs so that the
 NLO results for the larger 
 values of the BRs are quite similar. 
 
\section{Derivation of $\Delta v_{S}$}\label{ap:DvS}
   \begin{figure}[tbhp]     \label{fig:app2} 
   \begin{center}        
        \begin{minipage}{0.5\hsize}
        \begin{center}
        \includegraphics[width=80mm]{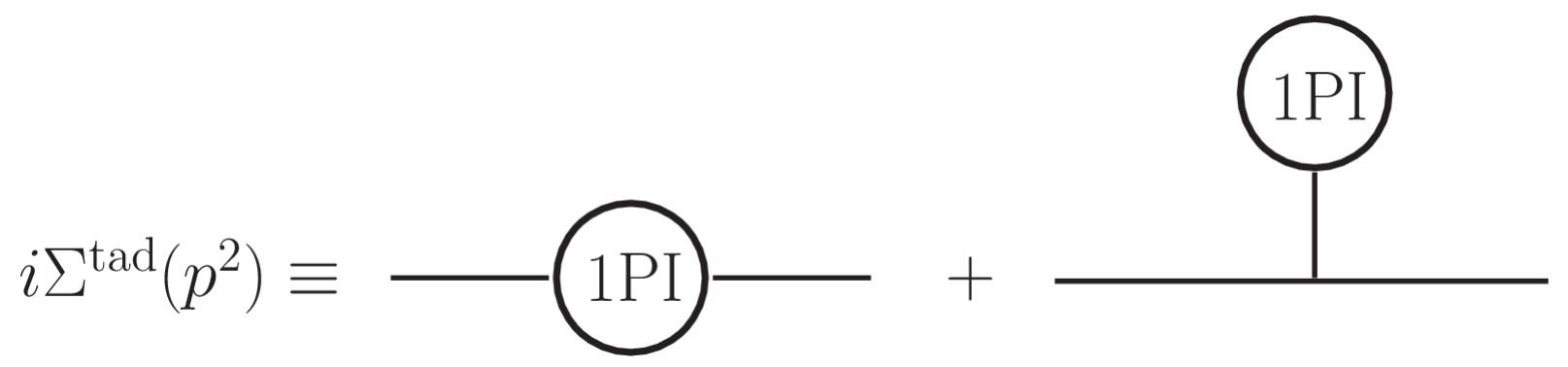}   
          \end{center}
                \end{minipage}\\
      \begin{minipage}{0.5\hsize}
     \begin{center}
        \includegraphics[width=80mm]{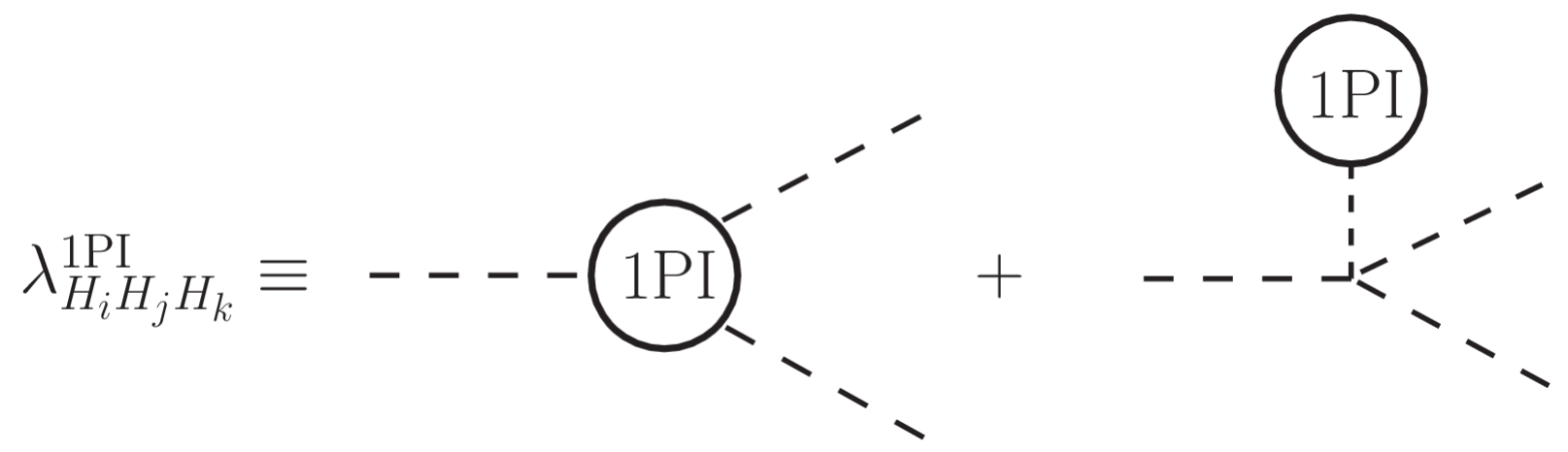}
              \end{center}
                \end{minipage}
     \vspace{0.5cm}
      \caption{ Self-energy diagrams and triangle vertex diagrams in the alternative tadpole scheme. 
      }
  
  \label{svtad}
      \end{center}  
\end{figure}
In this appendix, we derive the analytic expressions for $\Delta v_{S}$
for the case where ${ \lambda_{8}}$ and  ${m_{22}^{2}}$ are
renormalized in the $\overline{\rm MS}$ scheme. As mentioned before, if these 
parameters are renormalized via a physical process there in no need to
renormalize $v_S$. We stated in Sec.~\ref{sec:DvS}, that $\delta v_{S}$ is determined such that the remaining UV divergence in  the renormalized one-loop 
amplitude for $H_{1}\to H_{D} H_{D}$ is absorbed by the $\Delta v_{S}$
term in the process. \s

As schematically depicted in Fig.~\ref{svtad}, self-energies and one-loop amplitudes for $H_{1}\to H_{D}H_{D}$  can be separated into two parts: diagrams coming from the traditional tadpole scheme
and new diagrams including tadpole contributions due to the
alternative tadpole scheme. This is in fact the main difference
between the two schemes. 
Quantities in the usual tadpole scheme will be denoted by $X|_{\rm usual}$ while
tadpole contributions for the quantities $X$, which correspond to the second
diagrams in Fig.~\ref{svtad}, are written as $X|_{\rm tad}$.
One can check that, in the usual tadpole scheme, the UV divergences in $\amp^{\rm 1-loop}_{H_{1}\to H_{D}H_{D}}$ are cancelled without the need for introducing $\Delta v_{S}$. 
By using the ${\overline{\rm MS}}$ counterterms $\delta \lambda_{8}$ and $\delta m_{22}^{2}$, we can show that
\begin{align}
\left.\amp^{\rm 1PI}_{H_{1}\to H_{D}H_{D}}\right|_{\rm div, usual}
+\left. \amp^{\rm CT}_{H_{1}\to H_{D}H_{D}}\right|_{\rm div, usual}^{\Delta v_{S}=0}
=0. 
\end{align}
In the following paragraphs, we will show that this is not the case in
the alternative tadpole scheme. There are UV divergences coming from
the tadpole diagrams in the one-loop amplitude $\amp^{\rm
  1-loop}_{H_{1}\to H_{D}H_{D}}$ 
that lead to an extra infinity in the amplitude that will be cancelled by the $v_S$ counterterm.
First, we define the tadpole diagrams as
\begin{align}
T_{H_{i}}\equiv
\includegraphics[width=12mm]{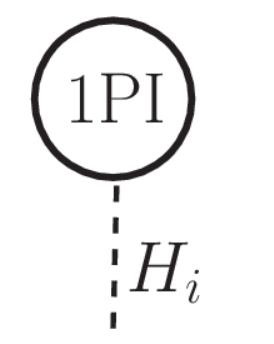}
\vspace{-2.0cm}
\end{align}
where $i=1,2$. Then the tadpole parts of  the 1PI diagram contributions are expressed as 
\begin{align}
\left.\amp^{\rm 1PI}_{H_{1}\to H_{D}H_{D}}\right|_{\rm tad}=\lambda_{H_{1}H_{1}H_{D}H_{D}}\frac{T_{H_{1}}}{m_{H_{1}}^{2}}+
\lambda_{H_{1}H_{2}H_{D}H_{D}}\frac{T_{H_{2}}}{m_{H_{2}}^{2}},
\end{align}
where
\begin{align}
\label{eq:lam11DD}
\lambda_{H_{1}H_{1}H_{D}H_{D}}&=\frac{c^{2}_{\alpha}}{v^{2}}(2m_{22}^{2}-2m_{H_{D}}^{2}+\lambda_{8}v_{S}^{2}) -s_{\alpha}^{2}\lambda_{8}, \\ 
\label{eq:lam12DD}
\lambda_{H_{1}H_{2}H_{D}H_{D}}&=
-\frac{c_{\alpha}s_{\alpha}}{v^{2}}\Big\{2m_{22}^{2}-2m_{H_{D}}^{2}+\lambda_{8}(v_{S}^{2}+v^{2})\Big\}. 
\end{align}
In the ${\overline{\rm MS}}$ scheme, the counterterms for $\delta \lambda_{8}$ and $\delta m_{22}^{2}$ do not contain tadpole contributions. The same 
is true  for  $\delta Z_{H_{D}H_{D}}$ and $\delta Z_{H_{1}H_{1}}$ because they are defined as the  derivatives of self-energies.
Therefore, they do not contribute to $\amp^{\rm CT}_{H_{1}\to H_{D}H_{D}}$, which allows us to write
\begin{align}
\notag
\left . \amp_{H_{1}\to H_{D}H_{D}}\right|^{\Delta v_{S}=0}_{\rm tad}&=
\lambda_{H_{2}H_{D}H_{D}}\left(\frac{1}{2}\delta Z_{H_{2} H_{1}}+\delta \alpha\right)\Bigg|_{\rm tad}
\\ 
&
-2\frac{c_{\alpha}}{v^{2}}\left(m_{22}^{2}-m_{H_{D}}^{2}+\frac{1}{2}v_{S}\lambda_{8}\right)\Delta v \big|_{\rm tad}
-2\frac{c_{\alpha}}{v}\delta m_{H_{D}}^{2} \big|_{\rm tad}. 
\end{align}
The various terms are given by the following expressions:

\begin{itemize}
\item
$\underline{\frac{1}{2}\delta Z_{H_{2} H_{1}}+\delta \alpha}$:

\noindent
  We can see that the tadpole parts are cancelled out: 
 \begin{align}
 \notag
\left (\frac{1}{2}\delta Z_{H_{2} H_{1}}+\delta \alpha\right)_{\rm tad}&=\frac{1}{2}\frac{1}{m_{H_{1}}^{2}-m_{H_{2}}^{2}}\left[\Sigma^{\rm tad}_{H_{1}H_{2}}(m_{H_{2}}^{2})-\Sigma^{\rm tad}_{H_{2}H_{1}}(m_{H_{1}}^{2})\right]\big|_{\rm tad} \\ 
 &=0, 
  \end{align}
  where we have used the following expressions for the tadpole parts of $\Sigma_{H_{1}H_{2}}$ and $\Sigma_{H_{2}H_{1}}$
  \begin{align}
  \Sigma^{\rm tad}_{H_{1}H_{2}}|_{\rm tad}=\Sigma^{\rm tad}_{H_{2}H_{1}}|_{\rm tad}=
  \lambda_{H_{1}H_{D}H_{D}}\frac{T_{H_{1}}}{m_{H_{1}}^{2}}+
  \lambda_{H_{2}H_{D}H_{D}}\frac{T_{H_{2}}}{m_{H_{2}}^{2}}. 
  \end{align}
  
  \item
  $\underline{\Delta v}$:
  
  \noindent
  The tadpole contributions of the gauge boson $(V=Z, W)$ self-energies are given by 
 \begin{align}
 \Sigma^{\rm tad}_{VV}|_{\rm tad}
 = -\left(c_{\alpha}\frac{T_{H_{1}}}{m_{H_{1}}^{2}}-s_{\alpha}\frac{T_{H_{2}}}{m_{H_{2}}^{2}}\right)\,.
 \end{align}
 This yields 
  \begin{align}
  \label{eq:dvevtad}
  \frac{\Delta v}{v}\Big|_{\rm tad}&=\frac{1}{2}\left(\frac{s_{W}^{2}-c_{W}^{2}}{s_{W}^{2}}\frac{1}{m^{2}_{W}}\Sigma_{WW}^{\rm tad}\big|_{\rm tad}+\frac{c_{W}^{2}}{s_{W}^{2}}\frac{1}{m^{2}_{Z}}\Sigma_{ZZ}^{\rm tad}\big|_{\rm tad}\right) \\ \notag
&=-\frac{1}{v}\left(c_{\alpha}\frac{T_{H_{1}}}{m_{H_{1}}^{2}}-s_{\alpha}\frac{T_{H_{2}}}{m_{H_{2}}^{2}}\right). 
  \end{align}

\item
$\underline{\delta m_{H_{D}}^{2}}$:
 
 \noindent 
The tadpole contribution for the mass counterterm $\delta m_{H_{D}}^{2}$ reads
  \begin{align}
  \label{eq:dmHDsqtad}
  \delta m_{H_{D}}^{2}|_{\rm tad}=\lambda_{H_{1}H_{D}H_{D}}\frac{T_{H_{1}}}{m_{H_{1}}^{2}}
  +\lambda_{H_{2}H_{D}H_{D}}\frac{T_{H_{2}}}{m_{H_{2}}^{2}}\,. 
  \end{align}
 
\end{itemize}
  
  Putting together all the results,  the tadpole part of $\amp^{\rm 1-loop}_{H_{1}\to H_{D}H_{D}}$can be written as
  \begin{align}\label{eq:amptad}
  \notag
  \amp^{\rm 1-loop}_{H_{1}\to H_{D}H_{D}}\big|_{\rm tad}^{\Delta v_{S}=0}&=
  \lambda_{H_{1}H_{1}H_{D}H_{D}}\frac{T_{H_{1}}}{m_{H_{1}}^{2}}+
\lambda_{H_{1}H_{2}H_{D}H_{D}}\frac{T_{H_{2}}}{m_{H_{2}}^{2}} \\ \notag
&+2\frac{c_{\alpha}}{v^{2}}\left(m_{22}^{2}-m_{H_{D}}^{2}+\frac{1}{2}v_{S}\lambda_{8}\right)\left(c_{\alpha}\frac{T_{H_{1}}}{m_{H_{1}}^{2}}-s_{\alpha}\frac{T_{H_{2}}}{m_{H_{2}}^{2}}\right) \\ \notag
&-2\frac{c_{\alpha}}{v}
\left(\lambda_{H_{1}H_{D}H_{D}}\frac{T_{H_{1}}}{m_{H_{1}}^{2}}
  +\lambda_{H_{2}H_{D}H_{D}}\frac{T_{H_{2}}}{m_{H_{2}}^{2}}\right) \\ \notag
  &=\frac{T_{H_{1}}}{m_{H_{1}}^{2}}\Big[
  \lambda_{H_{1}H_{1}H_{D}H_{D}}
  +2\frac{c_{\alpha}}{v^{2}}\left(m_{22}^{2}-m_{H_{D}}^{2}+\frac{1}{2}v_{S}\lambda_{8}\right)
  -2\frac{c_{\alpha}}{v}\lambda_{H_{1}H_{D}H_{D}}
  \Big] \\ 
  &+\frac{T_{H_{2}}}{m_{H_{2}}^{2}}\Big[
  \lambda_{H_{1}H_{2}H_{D}H_{D}}
  -2\frac{c_{\alpha}s_{\alpha}}{v^{2}}\left(m_{22}^{2}-m_{H_{D}}^{2}+\frac{1}{2}v_{S}\lambda_{8}\right)
  -2\frac{c_{\alpha}}{v}\lambda_{H_{2}H_{D}H_{D}}
  \Big]  \notag \\
   &=
{\lambda_{8}} \left(2\frac{{v_{S}}}{v} c_{\alpha }- s_{\alpha} \right) 
\delta v_{S}\,. 
  \end{align}
 In the last equality, we have used Eqs.~\eqref{eq:lam1DD},
 \eqref{eq:VEVCT},\eqref{eq:lam11DD} and \eqref{eq:lam12DD}. 
Because of $(\delta v_{S})_{\rm div}\neq 0$, 
 the UV divergence, which is proportional to $\lambda_{8}$, remains. 
Apart from this remaining term, we note that  terms with $m_{22}^{2}$ as well as $m_{H_{D}}^{2}$ are cancelled out. \s

The remaining UV-divergent term in Eq.~\eqref{eq:amptad} can be
absorbed by using the $\Delta {v_{S}}$ dependent part $\amp^{\rm 1-loop}_{H_{1}\to H_{D}H_{D}}\big|_{}^{\Delta v_{S}\neq0}$.  
Hence we set $\Delta v_{s}$ so as to eliminate the divergent part of Eq.~\eqref{eq:amptad},
\begin{align}
\label{eq:dvstad}
\Delta v_{S}=-(\delta v_{S})_{\rm div}. 
  \end{align}
Consequently, the one-loop amplitude for $H_{1}\to H_{D}H_{D}$ is UV finite.

\end{appendix}

\bigskip
\bigskip
\subsubsection*{Acknowledgments}
We thank Jorge Rom\~ao for fruitful discussions. DA, PG and RS are supported by FCT under contracts UIDB/00618/2020, UIDP/00618/2020, PTDC/FIS-PAR/31000/2017, CERN/FISPAR
/0002/2017, CERN/FIS-PAR/0014/2019, and by the HARMONIA project, 
contract UMO-2015/18/M/ST2/0518. The work of MM is supported by the
BMBF-Project 05H18VKCC1, project number 05H2018. 

\newpage

\end{document}